\documentclass[twoside,english]{elsarticle}
\usepackage[utf8]{inputenc}
\usepackage[T1]{fontenc}
\usepackage{geometry}
\usepackage{dsfont}
\usepackage{color} 
\usepackage{amsmath}
\usepackage{amssymb}
\usepackage{graphicx}
\usepackage{cancel}

\makeatletter
\journal{Journal of the Mechanics and Physics of Solids}

\usepackage{lineno} 
\usepackage{hyperref}
\usepackage[normalem]{ulem} 

\usepackage{array}
\usepackage{booktabs}
\usepackage{multirow}
\usepackage{rotating}

\makeatother

\usepackage{babel}

\usepackage{soul}
\usepackage[normalem]{ulem}
\usepackage{todonotes}

\begin{document}

\begin{frontmatter}


\title{The geometric nature of homeostatic stress in biological growth}

\author[LiPhy]{A.~Erlich\corref{cor1}}

\ead{alexander.erlich@univ-grenoble-alpes.fr}

\author[Galway]{G.~Zurlo\corref{cor1}}

\ead{giuseppe.zurlo@universityofgalway.ie}

\cortext[cor1]{Corresponding author}

\address[LiPhy]{Université Grenoble Alpes, CNRS, LIPHY, 38000 Grenoble, France}

\address[Galway]{School of Mathematical and Statistical Sciences, University of Galway, University Road, Galway, Ireland}

\begin{abstract}
 
Morphogenesis, the process of growth and shape formation in biological tissues, is driven by complex interactions between mechanical, biochemical, and genetic factors. Traditional models of biological growth often rely on the concept of homeostatic Eshelby stress, which defines an ideal target state for the growing body. Any local deviation from this state triggers growth and remodelling, aimed at restoring balance between mechanical forces and biological adaptation.  Despite its relevance in the biomechanical context, the nature of homeostatic stress remains elusive, with its value and spatial distribution often chosen arbitrarily, lacking a clear biological interpretation or understanding of its connection to the lower scales of the tissue. To bring clarity on the nature of homeostatic stress, we shift the focus from Eshelby stress to growth incompatibility, a measure of geometric frustration in the tissue that is the primary source of residual stresses in the developing body. Incompatibility, which is measured by the Ricci tensor of the growth metric at the continuous level, can be potentially regulated at the cell level through the formation of appropriate networks and connections with the surrounding cells, making it a more meaningful concept than homeostatic stress as a fixed target. In this geometric perspective, achieving a homeostatic state corresponds to the establishment of a physiological level of frustration in the body, a process leading to the generation and maintenance of the mechanical stresses that are crucial to tissue functionality. While residual stress can be induced through either active contraction or differential growth, the latter is the focus of this work. In this work we present a formulation of biological growth that penalises deviations from a desired state of incompatibility, similar to the way the Einstein-Hilbert action operates in General Relativity. The proposed framework offers a clear and physically grounded approach that elucidates the regulation of size and shape, while providing a means to link cellular and tissue scales in biological systems.

\end{abstract}

\begin{keyword} {morphogenesis; nonlinear elasticity;  growth laws; incompatibility; homeostatic stress; metric curvature;  variational methods. }
\end{keyword}

\end{frontmatter}{}

\setcounter{tocdepth}{2} 

\section{Introduction}

The determination of appropriate evolution equations for growth and shape change in morphogenesis has been a significant focus of research (e.g., \cite{epstein2000thermomechanics,lubarda2002mechanics,dicarlo2002growth,ambrosi2007growth,ganghoffer2010mechanical}). Due to the limited availability of experimental data, thermodynamic arguments are often employed to derive growth laws that satisfy a dissipation inequality. Typically, these models assume a single-constituent theory where each material point interacts with a mass reservoir at an imposed chemical potential, driving growth. The resulting growth law describes the evolution of growth as a response to stress stimuli, with growth ceasing once mechanical homeostasis is achieved \cite{epstein2000thermomechanics,dicarlo2002growth,ambrosi2007stress}:
\begin{equation}
T_{AB}=T_{AB}^{*}\,.\label{eq:mechanical-homeostasis-1}
\end{equation}
Here, $T_{AB}$ represents the Eshelby stress, and $T_{AB}^{*}$  is its homeostatic (target) value in physiological conditions. The indices $A$, $B$ refer to coordinates in the pre-grown reference configuration. The underlying hypothesis of mechanical homeostasis \cite{ambrosi2019growth,latorre2019mechanobiological,erlich2019homeostatic,goriely2017mathematics} suggests that growth and shape changes cease when (\ref{eq:mechanical-homeostasis-1}) is satisfied.

Residual stress, which persists in the tissue even in the absence of external loads or force fields, is ubiquitous in living tissues. It is observable when tissue integrity is disrupted, as seen in arteries \cite{chuong1986residual}, tumour spheroids \cite{stylianopoulos2012causes}, and the fruit fly wing disc \cite{harmansa2023growth}, among other biological systems \cite{Han1991,Gregersen2000,Omens2003}. When these tissues are radially cut, they open up, revealing the presence of residual stress. While residual stress can be induced through active contraction, our focus is on residual stress created through differential growth, where spatially varying growth creates geometrically incompatible stress-free states (internal frustration) in the body.  In this perspective, we intend to clarify how the prescription of a local balance of the Eshelby stress tensor as dictated by (\ref{eq:mechanical-homeostasis-1}) is linked to the onset of a spatially inhomogeneous state of residual stress in the body.

Despite their ubiquity in the biomechanical context, models relying upon (\ref{eq:mechanical-homeostasis-1}) raise an important issue concerning the choice of the homeostatic stress $T_{AB}^{*}$. A major issue concerns the arbitrary specification of $T_{AB}^{*}$: the up to six independent components of the homeostatic stress tensor $T_{AB}^{*}$, a spatially dependent field, must be provided by hand to calculate the equilibrium stress $T_{AB}$, yet these components are typically unknown and experimentally inaccessible. Furthermore, to create a plausible residual stress profile that corresponds well with experimental observations \cite{ambrosi2017solid,erlich2023mechanical,erlich2024incompatibility}, the target stress $T_{AB}^{*}$ must  be inhomogeneous and anisotropic. Providing the up to six independent components of an anisotropic, inhomogeneous tensor field $T_{AB}^{*}$ without direct experimental access necessitates considerable speculation.

To circumvent the somewhat arbitrary prescription of an anisotropic and inhomogeneous homeostatic stress $T_{AB}^{*}$,  multiphysics models with additional internal variables have been employed in literature.
Examples include multiphase or chemo-mechanical models in which a diffusing concentration field creates the spatial gradient that gives rise to residual stress, as commonly used in models of spheroids where oxygen diffuses through the tissue \cite{ambrosi2017solid,ambrosi2002mechanics}. In poroelastic models of growth, competition between a fluid and solid phase gives rise to residual stress \cite{ambrosi2002mechanics,xue2016biochemomechanical,fraldi2018cells}, whereas in mixture theories, cellular activity (e.g., fibroblasts or smooth muscle cells incorporating tensile stress into tissue fibers) creates residual stress \cite{humphrey2002constrained,cyron2017growth,ambrosi2019growth}.

To address difficulties arising from boundary conditions, some adaptations were introduced in literature. For example, $T_{AB}^{*}$ was chosen to be a dynamical variable that evolves in a delayed manner, trailing behind the evolution of $T_{AB}$, and thus reaching an oscillatory equilibrium rather than (\ref{eq:mechanical-homeostasis-1}). Alternatively, authors have chosen equilibrium relationships that differ from (\ref{eq:mechanical-homeostasis-1}), which however are not guaranteed to be consistent with the second law of thermodynamics \cite{erlich2019homeostatic}.

In this paper we propose a modification of the standard approach based on the target Eshelby stress, by explicitly accounting for the energetic cost of two physically distinct processes taking place in a growing biological body: the proper addition of mass (``growth'') and its internal organisation to produce incompatibility (``remodelling''), that is the only source of residual stresses in an isolated organ.  Inspired by the seminal work \cite{skalak1996compatibility}, where the relevance of compatibility was first recognized in the context of biological growth, and following preliminary results presented in \cite{erlich2024incompatibility}, we here quantify incompatibility by the Ricci curvature of the growth metric, and we claim that the homeostatic state is achieved when the following ``growth action'' is minimized: 
\begin{equation}
S=\int_{\mathcal{B}}\sqrt{G}\left((W-W^{*})+\frac{\lambda}{2}\left(\mathcal{R}-\mathcal{R}^{*}\right)^{2}\right)\,d^{3}X.
\label{eq:energy-functional}
\end{equation}
In this functional, $W^*$ is the chemical potential, characterising the nutrient availability in a surrounding nutrient bath, that is associated with growth, $\mathcal{R}$ is the Ricci scalar curvature associated with the growth metric tensor $G_{AB}$, $G$ is its determinant, and $\mathcal{R}^*$ is a target curvature representing the desired (homeostatic) level of geometric frustration in the growing body. $\mathcal{B}$ is a material manifold. The constant $\lambda$ is a coupling modulus quantifying the material's affinity to reach its prescribed target curvature. By incorporating both $W^*$ and $\mathcal{R}^*$ as main ingredients of the growth process, this formulation explicitly distinguishes between the energetic cost of adding mass and energetic cost of deviating from a physiological amount of frustration into the tissue. 

The main finding of this work is that the value of the homeostatic Eshelby stress arising from the minimisation of the growth action is
\begin{equation}
 T_{AB}^{*} = W^{*}G_{AB} + 2\lambda\left[\left(\mathcal{R} - \mathcal{R}^{*}\right)R_{AB} - \frac{1}{4}\left(\mathcal{R} - \mathcal{R}^{*}\right)^{2}G_{AB} - \nabla_{A}\nabla_{B}\mathcal{R} + G_{AB}\Delta\mathcal{R}\right]
 \label{eq:target-stress-quadratic}
\end{equation}
where \( \nabla_{A} \) is the covariant derivative and  \( \Delta = G^{AB} \nabla_{A} \nabla_{B} \) is the Laplace-Beltrami operator, and $R_{AB}$ is the Ricci tensor. This explicit expression of the target Eshelby stress in terms of chemical potential and curvature highlights the intricate mechanisms underlying the concept of homeostatic states in biology. From this mechanistic standpoint, biological tissues grow as long as there exists an imbalance between the energetic advantage of adding mass to the system, and the cost of storing elastically this mass in the body. Crucially, the storage of elastic energy necessitates growth to occur incompatibly, a process that may be locally regulated by targeting a physiological value of curvature.

In recent years, the formal study of evolution problems describing biological growth has gained interest \cite{davoli2023existence}. To the best of our knowledge, however, the investigation of minimisers for the growth action \eqref{eq:energy-functional} remains unexplored. Here, we demonstrate that, under specific symmetry conditions such as those governing growing discs and spheres, the homeostatic state corresponds to the minimisation of the growth action within the class of spherically symmetric solutions. This result extends the findings of \cite{erlich2024incompatibility}, which established that the equilibrium size of a growing organism is defined by $W^*$ and $\mathcal{R}^*$ for the specific geometry of a disc/sphere, and under the specific constraint $\mathcal{R}=\mathcal{R}^*$. In that setting, no general form of the homeostatic stress was derived and its geometric core remained hidden. The present work, recovering $\mathcal{R}=\mathcal{R}^*$ as a special case of a general theory, opens up a promising avenue in which incompatibility and its spatiotemporal dynamics and the thus generated opening patterns due to incisions can be theoretically and experimentally investigated. 
    
In practical terms, the proposed format simplifies the definition of a homeostatic state, firstly because only two scalar fields $W^*$ and $\mathcal{R}^*$ need to be prescribed, as opposed to the six independent components of the target homeostatic Eshelby stress; secondly, the targets $(W^*,\mathcal{R}^*)$ are not required to be inhomogeneous throughout the body, a necessity when defining the target Eshelby stress; lastly, the targets $W^*$ and $\mathcal{R}^*$ have a clear mechanical significance, as opposed to the components of the homeostatic Eshelby stress.
    
Several experimental observations supports the notion that incompatibility may be  uniformly distributed in real biological tissues, which is contained in the present model \eqref{eq:energy-functional} in the limit $\lambda\rightarrow\infty$. For instance, cutting experiments in \textit{Drosophila} wing discs reveal that radial incisions lead to a pronounced opening with outwardly curling edges (see, e.g., \cite{harmansa2023growth}), which is captured well by a growing disc with uniform positive curvature  ($\mathcal{R}=\mathcal{R}^*>0$) \cite[Fig. 1A]{erlich2024incompatibility}. The same model also captures the experimental observation that even repeated cuts of the wing disc do not fully relax the stress pattern. Similarly, cutting experiments of multicellular spheroids \cite{guillaume2019characterization}, \cite{stylianopoulos2012causes} reveal an opening pattern consistent with uniform positive Ricci curvature $\mathcal{R}=\mathcal{R}^*>0$ \cite[Fig. C]{erlich2024incompatibility}. Both the wing disc and multicellular spheroids exhibit growth up to a final size \cite[Fig. 1 B,D]{erlich2024incompatibility}, which once again is consistent with an evolution under uniform positive Ricci curvature. The present model \eqref{eq:energy-functional} expands this view by allowing for curvature variation $\mathcal{R}\neq\mathcal{R}^*$.

It should be noted that other models have considered opening patterns of multicellular spheroids, and some of these models indirectly relate size and residual stress accumulated due to differential growth. Ambrosi et al. \cite{ambrosi2017solid} considered a tumor growth model where the accumulation of residual stress is driven due to a chemical gradient and leads to an opening pattern with curled edges. In \cite{erlich2023mechanical} the anisotropy of a chemical potential tensor \cite{truskinovskiy1983chemical} creates residual stress and directly affects the final size of a growing spheroid. Other models for residual stress in spheroids were proposed, in which the buildup of residual stress is not explicitly modeled as a dynamic growth process. Differential growth between a growing tumor encapsulated by a shell in tension \cite{stylianopoulos2012causes} was proposed as a mechanism consistent with round outward-curling edges resulting from a radial cut. Explicit modelling of surface tension in a spheroid has a similar effect on the shape after cuts \cite{riccobelli_elastocapillarity-driven_nodate}.

While here we do not delve into the study of how incompatibility can be generated and controlled at lower scales, we note that the local regulation of curvature can be achieved, for instance, through deviations from target values of area and perimeter at the cell level or through local cell-to-cell rearrangements 
\cite{farhadifar2007influence, erlich2024incompatibility}. Also active contractility, ultimately resulting in stress generation and control at the scale of stress fibres or sarcomeres, may be viewed in principle as a process of active control of incompatibility, underscoring the broader applicability of the proposed theory also beyond differential growth.

    The work is organised as follows: after introducing the basic notions of Riemannian differential geometry and its connections to nonlinear elasticity in Sec.\ref{sec:Curved-space-theory}, we derive the homeostatic Eshelby stress as an Euler-Lagrange equation in Sec.\ref{subsec:derivation-of-homeostatic-stress}. We then discuss the energetic penalisation of curvature in Sec.\ref{sec:choice-of-coupling}, followed by an analysis of the kinematics, curvature, and mechanics of the residually stressed spheroid in Sec.\ref{sec:spheroid-kinematics}. The problem is greatly simplified by performing a gauge transformation taking into account internal symmetries of the system, as shown in Sec.  \ref{sec:gauge-fix}. The homeostatic equilibrium is examined in Sec.\ref{sec:homeostatic-equilibrium}, and we conclude with a discussion in Sec.\ref{sec:Discussion}.

\section{\label{sec:Curved-space-theory} An overview of residually stressed elasticity and its Riemannian interpretation}

The description of residual stresses in the context of biomechanics and growth mechanics is typically approached in two somewhat equivalent ways: the one based on the multiplicative decomposition of the deformation gradient \cite{KonadurovNikitin}, the other based on a Riemannian approach (e.g. \cite{eckart1948thermodynamics,marsden2012mathematical,yavari2010geometric}). In the multiplicative approach, the total deformation gradient $\mathbf{F}$ is decomposed into an elastic ($\mathbf{F}_{e}$) and anelastic ($\mathbf{F}_{g}$) part, $\mathbf{F}=\mathbf{F}_{e}\mathbf{F}_{g}$. A key distinction is whether the anelastic tensor, or in our context the growth tensor, $\mathbf{F}_{g}$ can be expressed as the gradient of a deformation map, or not. If it can, growth is called ``compatible'', and will not induce any residual stress, i.e. internal stress that remains in the absence of external fields or tractions on the boundary. In this case, the Riemannian curvature associated with the metric tensor $\mathbf{G}$ (which can be constructed as a quadratic form of $\mathbf{F}_{g}$) would be zero. The situation is different in the case of ``incompatible growth'' \cite{goriely2017mathematics}. In this case, $\mathbf{F}_{g}$ is not  a gradient field, and consequently the metric tensor $\mathbf{G}$ would produce a non-vanishing Riemann curvature tensor, leading to residual stresses in the body. This requires the introduction of an intermediate configuration, sometimes also called the ``virtual configuration'', which has no global isometric immersion into Euclidean space \cite{mcmahon2009geometry,mcmahon2011nonlinear} and is therefore often called ``non-Euclidean''. In practical calculations in the multiplicative decomposition framework, calculations are typically carried out in the initial (pre-growth) configuration, where Euclidean vectors can be easily defined. In this case, the Riemannian curvature due to the incompatibility of $\mathbf{F}_{g}$ is computed from the derived stress-free metric $\mathbf{G}$.

The Riemannian, or differential geometric approach to elasticity \cite{eckart1948thermodynamics,marsden2012mathematical,mcmahon2009geometry,goriely2017mathematics,yavari2010geometric}, to which we shall simply refer as the ``metric theory'', more clearly exposes the key role of curvature, which is crucial for the development of the present work. In this approach, the non-Euclidean nature of the virtual configuration, i.e. its curvature in the case of incompatible growth, is explicitly dealt with through the machinery of the differentiable manifold. When this manifold is paired with a Riemannian metric tensor, $G_{AB}$, the Riemannian manifold allows us to define vectors, tensors, differentiation and integration even if they do not happen in Euclidean space: the natural basis is then known as a coordinate basis, which forms a vector space in the tangent space $T_{\mathbf{X}}\mathcal{B}$ of a point $\mathbf{X}$ of the Riemannian manifold $T_{\mathbf{X}}\mathcal{B}$. In this framework, Ricci curvature is naturally defined as a contraction of the Riemann curvature tensor, which can be expressed through the metric $G_{AB}$ and its first and second spatial derivatives. 

\begin{figure}[t]
\hfill{}\includegraphics{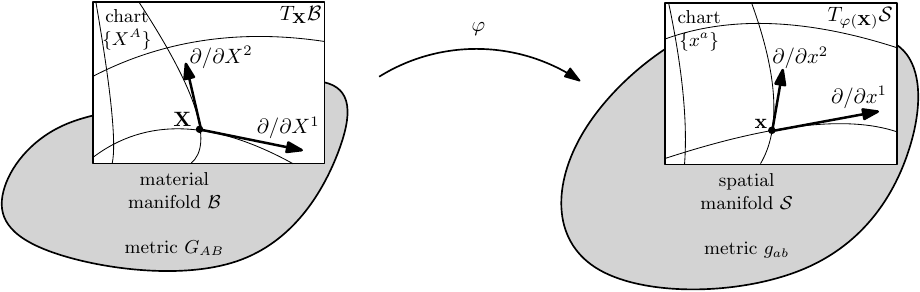}\hfill{} \caption{\label{fig:cartoon-potato-METRIC}Schematic of the  material manifold $\mathcal{B}$ (metric $G_{AB}$), which has curvature, and the flat spatial manifold $\mathcal{S}$ (metric $g_{ab}$). The deformation map $\varphi$ connects tangent spaces $T_{\mathbf{X}}\mathcal{B}$ and $T_{\varphi(\mathbf{X})}\mathcal{S}$. Their respective coordinate bases are illustrated.}
\label{fig:metric-theory-schematic}
\end{figure}

\subsubsection*{Deformation}

A body $\mathcal{B}$ is identified with a \textit{Riemannian manifold} $\mathcal{B}$ and a configuration of $\mathcal{B}$ is a mapping $\varphi:\mathcal{B}\rightarrow\mathcal{S}$, where $\mathcal{S}$ is another Riemannian manifold, see Fig. \ref{fig:metric-theory-schematic} . It is assumed that the body is stress-free in the material manifold. For a fixed $t$, $\varphi_{t}(\mathbf{X})=\varphi(\mathbf{X},t)$ and for a fixed $\mathbf{X}$, $\varphi_{\mathbf{X}}(t)=\varphi(\mathbf{X},t)$, where $\mathbf{X}$ is the position of material points in the undeformed configuration $\mathcal{B}$. Let $\varphi:\mathcal{B}\rightarrow\mathcal{S}$ be a $\mathcal{C}^{1}$ configuration of $\mathcal{B}$ in $\mathcal{S}$, where $\mathcal{B}$ and $\mathcal{S}$ are manifolds. The deformation gradient is the tangent map of $\varphi$ and is denoted by $\mathbf{F}=T\varphi$. Thus, at each point $\mathbf{X}\in\mathcal{B}$, it is a linear map 
\begin{equation}
\mathbf{F}(\mathbf{X}):T_{\mathbf{X}}\mathcal{B}\rightarrow T_{\varphi(\mathbf{X})}\mathcal{S}.
\end{equation}
We denote $\{X^{A}\}$ the local coordinate chart on $\mathcal{B}$, and the corresponding coordinate basis is $\{\partial_{A}\}$ and dual basis $dX^{A}$, so that $\partial_{A}\left(dX^{B}\right)=\delta_{A}^{B}$. Analogously, $\{x^{a}\}$ is the local chart on $\mathcal{S}$, with coordinate basis $\{\partial_{a}\}$, and dual basis $dx^{a}$, so that $\partial_{a}\left(dx^{b}\right)=\delta_{a}^{b}$. Throughout the text, we use latin uppercase letters $A$, $B$, $C$ etc. which run over three spatial dimensions to denote indices in the coordinate chart $\{X^{A}\}$. Similarly, lowercase indices $a$, $b$, $c$ etc. running over three spatial dimensions are used for the chart $\left\{ x^{a}\right\} $. This convention is used in the abbreviated notation $\partial_{A}=\partial/\partial X^{A}$ and $\partial_{a}=\partial/\partial x^{a}$. 

The components of $\mathbf{F}=F_{\;\:A}^{a}\partial_{a}\otimes dX^{A}$ are 
\[
F_{\;\:A}^{a}(\mathbf{X})=\partial_{A}\varphi^{a}(\mathbf{X})
\]

\subsubsection*{Metric tensors; Left and Right Cauchy Green tensors}

Here, $\mathbf{g}$ and $\mathbf{G}$ are metric tensors on $\mathcal{S}$ and $\mathcal{B}$, respectively. For brevity, we denote the determinant of $\mathbf{G}$ 
\begin{equation}
G=\det G_{AB}\,.
\end{equation}
In the geometric theory the following relation holds between volume elements of $(\mathcal{B},\mathbf{G})$ and $(\mathcal{S},\mathbf{g})$: 
\begin{equation}
\mathrm{d}v=J\,\mathrm{d}V
\end{equation}
where 
\begin{equation}
J=\sqrt{\frac{\det g_{ab}}{G}}\det F_{\;\:A}^{a}\,.\label{eq:J-metric-theory}
\end{equation}
Incompressibility of elastic deformations means that $J=1$. 

The left and right Cauchy Green tensor, $\mathbf{B}$ and $\mathbf{C}$, are \cite{marsden2012mathematical,yavari2010geometric}: 
\begin{equation}
B^{ab}=F_{\;\:A}^{a}G^{AB}\left(F^{T}\right)_{B}^{\;\:b}\qquad\text{and}\qquad C_{AB}=\left(F^{T}\right)_{A}^{\;\:a}g_{ab}F_{\;\:B}^{b}\,.\label{eq:Cauchy-green-tensors}
\end{equation}

\subsubsection*{Connection and covariant differentiation}

The connection coefficients $\Gamma_{IJ}^{K}$ of the Riemannian manifold $\left(\mathcal{B},\mathbf{G}\right)$ in the local chart $\{X^{A}\}$, and the connection coefficients $\gamma_{ij}^{k}$ of the Riemannian manifold $\left(\mathcal{S},\mathbf{g}\right)$ in the local chart $\{x^{a}\}$, are, respectively
\begin{equation}
\Gamma_{IJ}^{K}=\frac{1}{2}G^{KL}\left(\partial_{I}G_{JL}+\partial_{J}G_{IL}-\partial_{L}G_{IJ}\right),\qquad\gamma_{ij}^{k}=\frac{1}{2}g^{kl}\left(\partial_{l}g_{jl}+\partial_{j}g_{il}-\partial_{l}g_{ij}\right)\,.
\label{eq:Levi-Civita}
\end{equation}
The Levi-Civita connection, used throughout the text, is the unique torsion-free ($\Gamma_{IJ}^{K}=\Gamma_{JI}^{K}$, $\gamma_{ij}^{k}=\gamma_{ji}^{k}$) and metric-compatible ($\nabla_{A}G_{BC}=0$, $\nabla_{a}g_{bc}=0$) connection. 

Next, we give the expressions for covariant differentiation of second order tensors. First, consider some $\left(1,1\right)$ tensor $M_{\;\:B}^{A}$ that has both basis vectors in the material manifold $\left(\mathcal{B},\mathbf{G}\right)$, i.e. $\mathbf{M}=M_{\;\:B}^{A}\partial_{A}\otimes dX^{B}$. Similarly, consider some $\left(1,1\right)$ tensor $m_{\;\:b}^{a}$ that defined on $\left(\mathcal{S},\mathbf{g}\right)$, i.e. $\mathbf{m}=m_{\;\:b}^{a}\partial_{a}\otimes dx^{b}$. The covariant derivatives of $\mathbf{M}$ and $\mathbf{m}$ are 
\begin{equation}
\nabla_{C}M_{\;\:B}^{A}=\partial_{C}M_{\;\:B}^{A}+\Gamma_{CD}^{A}M_{\;\:B}^{D}-\Gamma_{CB}^{D}M_{\;\:D}^{A}\,,\qquad\nabla_{c}m_{\;\:b}^{a}=\partial_{c}m_{\;\:b}^{a}+\gamma_{cd}^{a}m_{\;\:b}^{d}-\gamma_{cb}^{d}m_{\;\:d}^{a}\,.\label{eq:covariant-derivative-definition}
\end{equation}
Covariant differentiation of a mixed-basis tensor, also called two-point tensor, such as the Piola-Kirchhoff stress $\mathbf{P}=P^{aA}\partial_{a}\otimes\partial_{A}$, requires a special covariant derivative that involves the connection coefficients in both manifolds $\mathcal{B}$ and $\mathcal{S}$ \cite{marsden2012mathematical}. The divergence of $\mathbf{P}$ is
\begin{equation}
\nabla_{A}P^{aA}=\frac{\partial P^{aA}}{\partial X^{A}}+\Gamma_{\;AB}^{A}P^{aB}+P^{bA}\gamma_{\;bc}^{a}F_{\;A}^{c}\,.\label{eq:covariant-derivative-two-point-tensor}
\end{equation}

\subsubsection*{Curvature}

The Riemann curvature tensor of the Riemannian manifold $\left(\mathcal{B},\mathbf{G}\right)$ in terms of the Levi-Civita connection is
\begin{equation}
R_{\;\:BCD}^{A}=\partial_{C}\Gamma_{BD}^{A}-\partial_{D}\Gamma_{BC}^{A}+\Gamma_{CE}^{A}\Gamma_{BD}^{E}-\Gamma_{DE}^{A}\Gamma_{BC}^{E}.\label{eq:Riemann-curvature-tensor}
\end{equation}
The Ricci curvature tensor $R_{AB}$, and its trace, Ricci scalar $\mathcal{R}$, are:
\begin{equation}
R_{AB}=R_{\;\:ACB}^{C},\qquad\text{\ensuremath{\mathcal{R}=G^{AB}R_{AB}=R_{\;\:A}^{A}.}}\label{eq:Ricci-tensor}
\end{equation}

The Riemannian manifold $\left(\mathcal{S},\mathbf{g}\right)$ is taken to be flat, i.e. free of curvature, so that the Riemann curvature tensor (expressed in terms of the Levi-Civita connection $\gamma_{ij}^{k}$) vanishes identically: $R_{\;\:bcd}^{a}=0$. Consequently, the Ricci tensor vanishes $R_{ab}=0$, as does the Ricci scalar $g^{ab}R_{ab}=R_{\;\:a}^{a}=0$.

\subsubsection*{Integration on manifolds; submanifolds; extrinsic curvature}

The natural volume element (also known as the volume form, or the Levi-Civita tensor) associated with a coordinate basis is 
\begin{equation}
dV=\sqrt{G}dX^{1}\wedge dX^{2}\wedge dX^{3}=\sqrt{G}d^{3}X\,,
\label{eq:volume-form}
\end{equation}
where $d^{3}X$ (or $d^{2}X$ on surfaces) is a shorthand for the wedge product.

Stokes’ theorem on a manifold relates the integral of the divergence of a field over a three-dimensional region $\mathcal{M}$ 
 to the flux of the field across the boundary $\partial\mathcal{M}$:
\begin{equation}
\int_{\mathcal{M}}d^{3}X\sqrt{G}\nabla_{A}V^{A}=\int_{\partial\mathcal{M}}d^{2}X\sqrt{H}N_{A}V^{A}\:.\label{eq:stokes-theorem}
\end{equation}
Here $V^{A}$ is some vector field in the material manifold. The normal outwards-pointing vector on $\partial\mathcal{M}$ is $\mathbf{N}=N^{A}\partial_{A}$. The induced metric $H_{AB}$ on the boundary, as well as its determinant are
\begin{equation}
H_{AB}=G_{AB}-N_{A}N_{B},\qquad H=\det H_{AB}\qquad\text{on }\partial\mathcal{B}\,.\label{eq:induced-metric}
\end{equation}
Furthermore, an important quantity on the boundary $\partial\mathcal{B}$ is the  extrinsic curvature 
\begin{equation}
K=\nabla_{A}N^{A}\qquad\text{on }\partial\mathcal{B}\,,\label{eq:extrinsic-curvature-scalar}
\end{equation}
which is the trace $K=K_{\;\:A}^{A}$ of the second fundamental form $K_{AB}$.


\section{\label{subsec:derivation-of-homeostatic-stress} Derivation of the homeostatic stress tensor}

\subsection{The energetic cost of maintaining a physiological target incompatibility}

Our theory is based on an energetic penalization of incompatibility, as measured by the Ricci curvature scalar~\(\mathcal{R}\), that can be accumulated in the body during growth. This energetic penalty measures the cost of deviating from an incompatible stress-free target state, in similar fashion to what happens in dislocation mechanics and plasticity with the energetic penalization of the dislocation density \cite{katanaev1992theory,Berdichevsky}. The specific functional dependence of the energy on curvature is not \emph{a priori} known. One may consider the following general form of an energy density
\begin{equation}
W+\lambda f(\mathcal{R})
\end{equation}
where \( W \) is the elastic strain energy density, and \( \lambda \) is the curvature modulus, a material parameter. In this formulation, \( \mathcal{R} \) denotes the Ricci scalar curvature derived from the growth metric tensor \( G_{AB} \) and serves as a quantitative measure of growth incompatibility, arising from spatially non-uniform growth. Incompatibility reflects the inherent challenge of assembling the grown parts without the necessity of creating voids or overlaps, effectively acting as the geometric "seed" of residual stress within the growing tissue. 
The function \( f(\mathcal{R}) \) encapsulates the dependence of the energy on curvature, and its specific form is not predetermined at this stage. For instance, the 
 linear coupling \( f(\mathcal{R}) = \mathcal{R} - \mathcal{R}^* \) is reminiscent of General Relativity, whereas \( f(\mathcal{R}) = (\mathcal{R} - \mathcal{R}^*)^2 \) provides a quadratic coupling with curvature that is more common in mechanics (e.g., plate theory). In the absence of a more direct knowledge, however,  for the time being we will leave \( f(\mathcal{R}) \) unspecified, and we will provide some hint on the possible functional dependence on curvature at a later stage. 

 \subsection{Momentum balance} 
 
 In the absence of external body force and inertia, we write momentum balance as 
 \begin{equation}
\nabla_{A}P_{a}^{\;\:A}=0\qquad\text{in }\mathcal{B}\,,\qquad\qquad P_{a}^{\;\:A}N_{A}=0\qquad\text{on }\partial\mathcal{B}\,.\label{eq:momentum-balance}
\end{equation}
where $P_{a}^{\;\:A}$ is the first Piola-Kirchhoff stress tensor. Here,  \( N_{A} \) is the outward unit normal vector on the boundary \( \partial \mathcal{B} \) and $\nabla_A$ a covariant derivative \eqref{eq:covariant-derivative-two-point-tensor}. The angular momentum balance imposes symmetry of the Cauchy stress tensor $\sigma^{ab}=\sigma^{ba}$, and the Cauchy and Piola stresses are related by $P_{a}^{\;\:A}=Jg_{ac}\sigma^{cb}\left(F^{-T}\right)_{b}^{\;\:A}
$ where the determinant of the deformation $J$ is defined in \eqref{eq:J-metric-theory}.

\subsection{First and second principle}
Combining the first and second principles of thermodynamics, at a fixed temperature, the dissipation $\Theta$ takes a simple form $\Theta=\dot{\mathcal{E}}+\dot{\mathcal{W}}-\dot{\mathcal{F}}\geq0$, where $\mathcal{E}$ is the free energy exchanged with the external environment (chemical potential), $\mathcal{W}$ is the  mechanical work performed by the environment on the tissue through its boundary, and $\mathcal{F}$ is the Helmholtz free energy of the tissue. We thus assume that mass is delivered within the bulk of the tissue, therefore:

\begin{equation}
\Theta=\underbrace{\frac{\mathrm{d}}{\mathrm{d}t}\int_{\mathcal{B}}\sqrt{G}\,W^{*}\,d^{3}X}_{\dot{\mathcal{E}}}+\underbrace{\int_{\mathcal{B}}\sqrt{G}\,P_{\;\:a}^{A}\nabla_{A}\dot{\varphi}^{a}\,d^{3}X}_{\dot{\mathcal{W}}}-\underbrace{\frac{\mathrm{d}}{\mathrm{d}t}\int_{\mathcal{B}}\sqrt{G}\left(W+\lambda f\left(\mathcal{R}\right)\right)\,d^{3}X}_{\dot{\mathcal{F}}}\geq0\,.\label{eq:general-dissipation}
\end{equation}
The dissipation rate \(\Theta\) in equation \eqref{eq:general-dissipation} quantifies the irreversible entropy production within the growing tissue and must satisfy \(\Theta \geq 0\) to comply with the second law of thermodynamics. Specifically, \(\dot{\mathcal{E}}\) represents the rate of free energy received from a surrounding nutrient bath with the prescribed chemical potential \(W^{*}\), see e.g.\cite{erlich2024incompatibility,buskohl2014influence} for an analogous term in the context of biological growth. The term \(\dot{\mathcal{W}}\) corresponds to the rate of external mechanical work performed on the tissue through traction forces, encapsulated by the first Piola-Kirchhoff stress tensor \(P_{\;\:a}^{A}\) and the gradient of the velocity field \(\nabla_{A}\dot{\varphi}^{a}\). Lastly, \(\dot{\mathcal{F}}\) denotes the rate of change of the tissue's Helmholtz free energy, which includes both the elastic strain energy density \(W\) and the energetic cost per unit current volume \(\lambda f\left(\mathcal{R}\right)\) which penalises a deviation from a target (physiological)  incompatibility, e.g. $f(\mathcal{R})=\frac{1}{2}(\mathcal{R}-\mathcal{R}^*)^2$. Combining these terms, the inequality \(\Theta = \dot{\mathcal{E}} + \dot{\mathcal{W}} - \dot{\mathcal{F}} \geq 0\) enforces non-negative entropy production.

In the study of growth, it is common to use the dissipation inequality \eqref{eq:general-dissipation} to derive a local form of the evolution law for the unknown stress-free metric $G_{AB}$, and we shall proceed this way.  For convenience we set 
\begin{equation}
S_{\text{M}} = \int_{\mathcal{B}}\sqrt{G}\,W\,d^{3}X, \qquad
S_{\text{C}} = \int_{\mathcal{B}}\sqrt{G}\, f(\mathcal{R})\,d^{3}X, \qquad
S_{\text{E}} = \int_{\mathcal{B}}\sqrt{G}\,W^* \,d^{3}X, \qquad
\label{eq:action-shorthands}
\end{equation}
for the {\it mechanical}, for the {\it curvature} and for the {\it external} contributions dissipation inequality. With these, the dissipation inequality \eqref{eq:general-dissipation} can be compactly restated 
\begin{equation}
\Theta=\dot{\mathcal{W}}-\frac{\mathrm{d}}{\mathrm{d}t}\left(S_{\text{M}}-S_{\text{E}}+S_{\text{C}}\right)\geq0\,.\label{eq:simplified-dissipation}
\end{equation}
In the following, we show how to take the time derivatives of each of these terms explicitly. For time derivatives we use the notation $\frac{\mathrm{d}}{\mathrm{d}t}$ and the overdot interchangeably. 

\subsection*{Time derivatives of the mechanical  $S_{\text{M}}$ and external $S_{\text{E}}$ terms}

The time derivatives $\dot{S}_{\text{M}}-\dot{S}_{\text{E}}$ read explicitly
\begin{equation}
\dot{S}_{\text{M}}-\dot{S}_{\text{E}} = \int_{\mathcal{B}} \left( W \, \frac{\mathrm{d}}{\mathrm{d}t} \sqrt{G} + \sqrt{G} \, \frac{\mathrm{d}}{\mathrm{d}t} W - W^* \, \frac{\mathrm{d}}{\mathrm{d}t} \sqrt{G} \right) d^{3}X,
\end{equation}
where \( \frac{\mathrm{d}}{\mathrm{d}t} \sqrt{G} = \frac{1}{2} \sqrt{G} \, G^{AB}  \dot{G}_{AB} \). The strain energy \( W \) is a function of the deformation gradient \( F_{\;\:A}^{a} \) and the metric tensor \( G^{AB} \), through a dependence on the  left Cauchy-Green tensor \( B^{ab} = F_{\;\:A}^{a} G^{AB} (F^{T})_{B}^{\;\:b} \), that is 
\begin{equation}
W=W\left(B^{ab}\right)\,.
\label{eq:material-isotropy}
\end{equation}
By this choice the material is assumed isotropic; the expression of the kinematic tensors in terms of quadratic forms $B^{ab}$ and $G_{AB}$ simplifies the analysis. With this, we obtain
\begin{equation}
\dot{S}_{\text{M}}-\dot{S}_{\text{E}}=\int_{\mathcal{B}}\sqrt{G}\,\left(2\frac{\partial W}{\partial B^{ab}}F_{\;\:B}^{b}G^{BA}\right)\nabla_{A}\dot{\varphi}^{a}\,d^{3}X-\frac{1}{2}\int_{\mathcal{B}}\sqrt{G}\,\left(T_{AB}-W^{*}G_{AB}\right)\dot{G}^{AB}\,d^{3}X,
\label{eq:variation-of-mechanical-action}
\end{equation}
The second term in \eqref{eq:variation-of-mechanical-action} involves the chemical potential $W^*$ and the Eshelby-like stress
\begin{equation}
T_{\;\:B}^{C}=W\delta_{B}^{C}-2G^{CA}(F^{T})_{A}^{\;\:a}\frac{\partial W}{\partial B^{ab}}F_{\;\:B}^{b}.
\label{eq:eshelby-def}
\end{equation}

\subsubsection*{Time derivative of the curvature term $S_{\text{C}}$}

The term \( S_{\text{C}} \) involves the Ricci scalar curvature \( \mathcal{R} \) and is given by \eqref{eq:action-shorthands}.
When taking the time derivative with respect to the inverse metric tensor \( G^{AB} \), we obtain both bulk and boundary term, $ \dot{S}_{\text{C}} =   \dot{S}_{\text{C}}^{\text{bulk}} +  \dot{S}_{\text{C}}^{\text{bnd}}$. The former arises from integration over the volume \( \mathcal{B} \) and includes contributions from the Ricci curvature and its derivatives. The variation of the Ricci curvature scalar is \cite{guarnizo2010boundary}:
\begin{equation}
\dot{\mathcal{R}}=R_{AB}\dot{G}^{AB}+G_{AB}\Delta(\dot{G}^{AB})-\nabla_{A}\nabla_{B}(\dot{G}^{AB})\,.
\end{equation}
Integrating the second gradient terms by parts twice using the Stokes theorem \eqref{eq:stokes-theorem}, the variation $S_\text{C}$ with respect to \( G^{AB} \) yields a well-known result in generalised theories of gravity \cite{sotiriou2010f}:
\begin{equation}
    \dot{S}_{\text{C}}^{\text{bulk}} = \int_{\mathcal{B}} \sqrt{G} \left[ f'(\mathcal{R}) R_{AB} - \tfrac{1}{2} f(\mathcal{R}) G_{AB} - \nabla_{A} \nabla_{B} f'(\mathcal{R}) + G_{AB} \Delta f'(\mathcal{R}) \right] \dot{G}^{AB} d^{3}X\,.
    \label{eq:variation-of-metric-action-BULK}
\end{equation}
Here, \( f'(\mathcal{R}) \) is the derivative of \( f(\mathcal{R}) \) with respect to \( \mathcal{R} \), 
\( \nabla_{A} \) denotes the covariant derivative with respect to the coordinate \( X^{A} \), and  \( \Delta = G^{AB} \nabla_{A} \nabla_{B} \) is the Laplace-Beltrami operator.

The boundary term arises due to integration by parts when dealing with the second-order spatial derivatives of \( \dot{G}^{AB} \), and is given by \cite[Eq. (59)]{guarnizo2010boundary},\cite[Eq. (3.17)]{sotiriou2007modified}
\begin{equation}
\dot{S}_{\text{C}}^{\text{bnd}}=-\int_{\partial\mathcal{B}}\sqrt{H}N^{C}M_{C}d^{2}X\,-\int_{\partial\mathcal{B}}\,\sqrt{H}N_{D}V^{D}d^{2}X\,,
    \label{eq:variation-of-metric-action-BOUNDARY}
\end{equation}
where \( \partial \mathcal{B} \) is the boundary of the volume \( \mathcal{B} \),  \( \sqrt{H} \) is the square root of the determinant of the induced metric \( H_{AB} \) on the boundary, see \eqref{eq:induced-metric}, and  \( K \) is the trace of the extrinsic curvature (mean curvature) \eqref{eq:extrinsic-curvature-scalar} of the boundary surface \( \partial \mathcal{B} \) with normal $N_A$. Finally, \( \sqrt{H} d^{2}X \) is the volume form for integration over the two-dimensional boundary surface \eqref{eq:volume-form},\eqref{eq:stokes-theorem}. The expressions $M_C$ and $V^D$ are given by
\begin{equation}
\begin{alignedat}{1}\begin{alignedat}{1}M_{C} & =\end{alignedat}
 & -f'(\mathcal{R})G^{AB}\nabla_{C}(\dot{G}_{AB})+G^{AB}\dot{G}_{AB}\nabla_{C}(f'(\mathcal{R}))\\
V^{D}= & -f'(\mathcal{R})G^{DA}G^{BC}\nabla_{B}(\dot{G}_{AC})+G^{DA}G^{BC}\dot{G}_{AC}\nabla_{B}(f'(\mathcal{R}))\,.
\end{alignedat}
\label{eq:boundary-subexpressions}
\end{equation}

\subsubsection*{Deriving constitutive laws from the dissipation inequality}

We now revisit the dissipation inequality \eqref{eq:simplified-dissipation}, with the mechanical and external terms $\dot{S}_{\text{M}}-\dot{S}_{\text{E}}$  given by \eqref{eq:variation-of-mechanical-action} and the curvature term $\dot{S}_{\text{C}}$ given as the sum of the bulk contribution \eqref{eq:variation-of-metric-action-BULK} and boundary contribution \eqref{eq:variation-of-metric-action-BOUNDARY}. 
The variation of the full growth action finally leads to 
\begin{equation}
\begin{alignedat}{1}\Theta & =\int_{\mathcal{B}}\sqrt{G}\,\left(P_{\;\:a}^{A}-2\frac{\partial W}{\partial B^{ab}}F_{\;\:B}^{b}G^{BA}\right)\nabla_{A}\dot{\varphi}^{a}\,d^{3}X+\frac{1}{2}\int_{\mathcal{B}}\sqrt{G}\,\left(T_{AB}-T_{AB}^{*}\right)\dot{G}^{AB}\,d^{3}X\\
 & +\int_{\partial\mathcal{B}}\sqrt{H}\left(N^{C}M_{C}+N_{D}V^{D}\right)d^{2}X\geq0
\end{alignedat}
\label{eq:full-variation-METRIC}
\end{equation}
where we have introduced the main character of this work, the homeostatic Eshelby stress
\begin{equation}
T_{AB}^{*} = W^{*} G_{AB} + 2\lambda \left( f'(\mathcal{R}) R_{AB} - \frac{1}{2} f(\mathcal{R}) G_{AB} - \nabla_{A} \nabla_{B} f'(\mathcal{R}) + G_{AB} \Delta f'(\mathcal{R}) \right).
\label{eq:homeostatic-equilibrium}
\end{equation}
Before drawing conclusions on its physical meaning in the context of growth, we proceed towards the derivation of constitutive laws from the dissipation inequality \eqref{eq:homeostatic-equilibrium}. We assume that  the velocity gradient \(\nabla_A\dot{\varphi}^{a}\), the stress-free metric \( \dot{G}_{AB}\) are also its spatial gradient\( \nabla_C\dot{G}_{AB}\)   are  varied independently. 

From the first integral in \eqref{eq:full-variation-METRIC}, 
we deduce
\begin{equation}
P_{a}^{\;\:A} = 2 \frac{\partial W}{\partial B^{ab}} F_{\;\:B}^{b} G^{BA}\,, \label{eq:piola-def}
\end{equation}
where $P_{a}^{\;\:A}$ is the first Piola-Kirchhoff stress tensor. Note that, using  the definition of the  Cauchy stress tensor \( \sigma_{aj} \), the Eshelby stress tensor $T_{\;\:B}^{C}$  \eqref{eq:eshelby-def} can be reformulated in the following useful way:
\begin{equation}
\sigma_{aj}=\frac{2}{J}\frac{\partial W}{\partial B^{ab}}B^{bk}g_{kj},\qquad\qquad T_{\;\:B}^{C}=W\delta_{B}^{C}-J\left(F^{-1}\right)_{\;\:b}^{C}\sigma^{bc}g_{cd}F_{\;\:B}^{d}\,,
\label{eq:Eshelby-definition}
\end{equation}
where \( J \) is the determinant of the deformation gradient \( F_{\;\:A}^{a} \), and \( g_{kj} \) is the spatial metric tensor.

The second term in \eqref{eq:full-variation-METRIC} describes the dynamics of the growth metric tensor \(G^{AB}\) in the bulk. The term $\left(T_{AB}-T_{AB}^{*}\right)\dot{G}^{AB}$ is dissipative. To ensure that the dissipation inequality is satisfied, we assume the close-to-equilibrium relation
\begin{equation}
\dot{G}_{AB}=k\left(T_{AB}^{*}-T_{AB}\right)
\label{eq:generic-growth-law}
\end{equation}
which we call a \textit{growth law}. Here, $k$ is a positive scalar parameter. Note that because $T_{AB}^{*}$ is symmetric, due to the symmetry of the Ricci and of the metric tensors, and due to the properties of the Levi-Civita connection, also $T_{AB}$ must be symmetric at equilibrium.  From here it follows  that the material must be of coaxial type
\begin{equation}
B^{ac}\sigma_{c}^{\;\:b}=\sigma^{ac}B_{c}^{\;\:b}\,.
\end{equation} 
which is consistent with  the assumption of material isotropy \eqref{eq:material-isotropy}  because a  hyperelastic material is coaxial if and only if it is isotropic \cite{Vianello1996}. 

We also note that the target stress tensor is divergence-free ($\nabla^{B}T_{AB}^{*}=0$)
as can be shown by a standard calculation involving the contracted
Bianchi identities and the Ricci identity. Furthermore, the energy-momentum
tensor is also divergence free $\nabla^{B}T_{AB}=0$, assuming that
linear momentum balance $\nabla_{A}P^{aA}=0$ and the constitutive
relationship \eqref{eq:piola-def} holds.  Thus, $T_{AB}^{*}$ satisfies
momentum balance by construction.

Finally, the third term in  \eqref{eq:full-variation-METRIC} is a boundary term arising from the variation of the growth metric tensor on the boundary. Noting that $\nabla_{C}f'\left(\mathcal{R}\right)=f''\left(\mathcal{R}\right)\partial_{C}\mathcal{R}$, we obtain the requirement 
\begin{equation}
f'(\mathcal{R})=0\quad\text{and}\quad f''\left(\mathcal{R}\right)\partial_{C}\mathcal{R}=0\quad\text{on }\partial\mathcal{B}\,.
\label{Rbnd}
\end{equation}
In the absence of specific physical mechanisms justifying an interaction between the external environment and curvature through the body's boundary, the condition $f'(\mathcal{R}) = 0$ should be interpreted as a natural outcome of the variational framework. More complex boundary conditions that generalize \eqref{Rbnd} arise in contexts such as surface growth, where the deposition of material on the boundary allows for control over the boundary value of incompatibility, by prescribing the full stress tensor and not merely tractions, as it happens in non-growing bodies \cite{ZTPRL,ZTMRC,TZPRE}. However, because surface growth is not considered here, we disregard the influence of the environment on the body's microstructure via its boundary, and we adopt \eqref{Rbnd} as a natural condition. \\

The full growth system, comprising mechanics and growth, is composed of 30 equations and  30 unknowns, summarised in Table  \eqref{tab:unknowns-and-equations}. While it is beyond the scope of this work to make any statements of the general well-posedness of the problem stated in Table \eqref{tab:unknowns-and-equations}, we will demonstrate in the following the successful solution of this system in spherical symmetry. 

\begin{table}[t]
\centering{%
\begin{tabular}{>{\centering}p{3cm}>{\centering}p{1.3cm}>{\centering}p{0.8cm}>{\centering}p{1cm}>{\centering}p{3cm}>{\centering}p{4.7cm}>{\centering}p{1.5cm}}
\toprule 
\textbf{Unknown} & \textbf{Symbol} & \textbf{Number} &  & \textbf{Nature} & \textbf{Equation} & \textbf{Number}\tabularnewline
\midrule 
deformation map & $\varphi^{a}$ & 3 &  & total deformation & $F_{\;\:A}^{a}=\partial_{A}\varphi^{a}$ & 9\tabularnewline
\midrule 
total deformation & $F_{\;\:A}^{a}$ & 9 &  & morphoelastic decomposition & $B^{ab}=F_{\;\:A}^{a}G^{AB}\left(F^{T}\right)_{B}^{\;\:b}$ & 6\tabularnewline
\midrule 
elastic def. & $B^{ab}$ & 6 &  & growth law & $\dot{G}_{AB}=k(T_{AB}^{*}-T_{AB})$ & 6\tabularnewline
 &  &  &  & boundary condition & $f'(\mathcal{R})=f''(\mathcal{R})\partial_C \mathcal{R}=0$ on $\partial\mathcal{B}$ & \tabularnewline
\midrule 
growth metric & $G_{AB}$ & 6 &  & constitutive law & $\sigma_{aj}=\frac{2}{J}\frac{\partial W}{\partial B^{ab}}B^{bk}g_{kj}$ & 6\tabularnewline
\midrule 
Cauchy stress & $\sigma_{ab}$ & 6 &  & momentum balance & $\nabla_{A}P^{aA}=0$ & 3\tabularnewline
 &  &  &  & boundary condition & $P_{\;\:A}^{a}N^{A}=0$ on $\partial\mathcal{B}$& \tabularnewline
\midrule 
 &  &  &  &  &  & \tabularnewline
 & \textbf{total} & \textbf{30} &  &  & \textbf{total} & \textbf{30}\tabularnewline
\bottomrule
\end{tabular}}\caption{\label{tab:unknowns-and-equations}Unknowns and equations for the dynamical problem. }
\end{table} 

\section{\label{sec:choice-of-coupling}Considerations on the energetic penalisation of curvature}

In the absence of a tissue-specific micro-mechanical model for the establishment of incompatibility, we adopt a mathematically driven approach to identify the simplest yet effective coupling between growth and curvature. As shown below, the simplest possible choice of linear penalisation of curvature (as done in General Relativity through the Einstein-Hilbert action) is not a good choice in the growth context, because it leads to unacceptable boundary conditions. This shortcoming can be addressed by a quadratic penalisation of curvature. 

\subsection{Why linear coupling with curvature must be discarded}

The most straightforward choice for the coupling function is a linear relationship of the form:
\begin{equation}
f(\mathcal{R}) = \mathcal{R} - \frac{1}{3}\mathcal{R}^*
\label{eq:linear-coupling}
\end{equation}
where \( \mathcal{R} \) is the Ricci scalar curvature and \( \mathcal{R}^* \) represents a target curvature. This linear coupling is appealing for a number of reasons. The action  $S_\text{C}$ becomes the celebrated Einstein-Hilbert action 
\begin{equation}
S_{\text{C}}=\int_{\mathcal{B}}\sqrt{G}\left(\mathcal{R}-\frac{1}{3}\mathcal{R}^{*}\right)\,d^{3}X
\label{eq:einstein-hilbert-action}
\end{equation}
Its variation, taken by itself ($\dot{S}_\text{C}=0$), leads to a minimum at $\mathcal{R}=\mathcal{R}^*$. The resulting equilibrium equations $T_{AB} = T_{AB}^*$ are formally analogous to the field equations of General Relativity \cite{wald2010general,carroll2004introduction}:
\begin{equation}
R_{AB}-\frac{1}{2}\mathcal{R}G_{AB}+\Lambda G_{AB}=\frac{1}{2\lambda}T_{AB}\qquad\text{in }\mathcal{B}\,,
\label{eq:Einsteins-equations}
\end{equation}
where we introduced a scalar constant $\Lambda$, that in General Relativity is known as the Cosmological Constant, and that here is given by $\Lambda=\frac{1}{2}\left(\frac{W^{*}}{\lambda}+\frac{1}{3}\mathcal{R}^{*}\right)$.

In General Relativity, $T_{AB}$ featured in  \eqref{eq:Einsteins-equations} is called the energy-momentum tensor, which is precisely the term used by J.D. Eshelby to name what we currently call ``Eshelby stress''  \cite{eshelby1975elastic}.


However, for how clean and suggestive \eqref{eq:Einsteins-equations} is, it presents significant challenges that make it unsuitable for our biological growth setting. Indeed, firstly note that the variation of the Einstein-Hilbert action \eqref{eq:einstein-hilbert-action} leads to an ill-posed variational problem \cite{chakraborty2017boundary}. This drawback is  well known in General Relativity, where the action is augmented by the Gibbons-Hawking-York (GHY) boundary term \cite{gibbons1977action,wald2010general,poisson2004relativist}:
\begin{equation}
S'_{\text{C}} = S_{\text{C}} + S_{\text{GHY}},
\end{equation}
where
\begin{equation}
S_{\text{GHY}} = 2 \int_{\partial \mathcal{B}} \sqrt{H} \, K \, d^2X.
\end{equation}
Here, \( H \) is the determinant of the induced metric on the boundary \( \partial \mathcal{B} \), and \( K \)   is the mean curvature scalar of the boundary surface \( \partial \mathcal{B} \), i.e. the trace of the second fundamental form.  The variation of the augmented action $\dot{S}'_\text{C}=0$ recovers the Einstein equations \eqref{eq:Einsteins-equations} in the bulk and renders the variational problem well-posed.

While the inclusion of the GHY term successfully regularises the variational problem of General Relativity in ensuring a well-posed boundary value problem, this approach does not seamlessly translate to  biological growth, where it is not possible to give a clear mechanical interpretation of the GHY term.

\subsection{Energy Density Couples Quadratically with Curvature}

A quadratic coupling of the form
\begin{equation}
f\left(\mathcal{R}\right) = \frac{1}{2}\left(\mathcal{R} - \mathcal{R}^{*}\right)^{2}
\label{eq:quadratic-coupling}
\end{equation}
retains the spirit of the Einstein-Hilbert action \eqref{eq:einstein-hilbert-action}  by ensuring that its minimisation is achieved for \( \mathcal{R} = \mathcal{R}^{*} \) (see also \cite{salvio2018quadratic}). 
    
In this case, the homeostatic equilibrium stress tensor \(  T_{AB}^{*} \) becomes the same as in the introduction \eqref{eq:target-stress-quadratic}:
\[
T_{AB}^{*}=W^{*}G_{AB}+2\lambda\left[\left(\mathcal{R}-\mathcal{R}^{*}\right)R_{AB}-\frac{1}{4}\left(\mathcal{R}-\mathcal{R}^{*}\right)^{2}G_{AB}-\nabla_{A}\nabla_{B}\mathcal{R}+G_{AB}\Delta\mathcal{R}\right]\qquad\text{in }\mathcal{B}\,.
\]
We will explore this form of the homeostatic stress tensor in the subsequent sections.\\

\section{\label{sec:spheroid-kinematics}Kinematics, curvature and mechanics of the residually stressed spheroid}

To show the predictive power of the theory, we now discuss the analytically manageable and physically transparent case of a growing spheroid. This study encompasses  the derivation of the corresponding equilibrium equations, the analysis of boundary conditions, and the examination of how the quadratic coupling influences the residual stress distribution in the growing spheroid. 

\subsection{\label{subsec:spheroid-kinematics}Kinematics}

We consider here an incompressible neo-Hookean spheroid that grows. Note that we denote with a prime the derivative with respect to the material radial coordinate $R$, that is $(\cdot)'=\partial(\cdot)/\partial R$. Also note our choice of coordinates $\Theta=\theta$, $\Phi=\phi$, although we sometimes keep distinguishing between the material and spatial angular coordinates for clarity. 

For a growing sphere, the metric tensor in the material manifold is

\begin{equation}
G_{AB}=\underbrace{\begin{pmatrix}\frac{1}{\Gamma\left(R\right)} & 0 & 0\\
0 & 1 & 0\\
0 & 0 & 1
\end{pmatrix}}_{\text{incompatible}}\underbrace{\left(\begin{array}{ccc}
g'(R)^{2} & 0 & 0\\
0 & g(R)^{2} & 0\\
0 & 0 & g(R)^{2}\sin^{2}\Theta
\end{array}\right)}_{\text{compatible}},\qquad G^{AB}=\left(\begin{array}{ccc}
\frac{\Gamma\left(R\right)}{g'(R)^{2}} & 0 & 0\\
0 & \frac{1}{g(R)^{2}} & 0\\
0 & 0 & \frac{1}{g(R)^{2}\sin^{2}\Theta}
\end{array}\right)\label{eq:metric-sphere-G}
\end{equation}
Here, as previously \cite{erlich2024incompatibility}, we use a split into compatible and incompatible part that greatly simplifies the notation: For the sake of clear notation, in what follows we use the variables $g\left(R\right)$ and $\Gamma\left(R\right)$. Note that $g$ is not the spatial metric determinant (which is denoted $\det g_{ab}$, see \eqref{eq:J-metric-theory}), and $\Gamma$ is not the material Christoffel symbol (denoted $\Gamma^{K}_{IJ}$, see \eqref{eq:Levi-Civita}).

The metric tensor in the spatial manifold is 
\begin{equation}
g_{ab}=\left(\begin{array}{ccc}
1 & 0 & 0\\
0 & r^{2} & 0\\
0 & 0 & r^{2}\sin^{2}\theta
\end{array}\right),\qquad g^{ab}=\left(\begin{array}{ccc}
1 & 0 & 0\\
0 & \frac{1}{r^{2}} & 0\\
0 & 0 & \frac{1}{r^{2}\sin^{2}\theta}
\end{array}\right)\,.\label{eq:sphere-g}
\end{equation}
The square roots of the determinants of the material and spatial metric tensors are 
\begin{equation}
\sqrt{G}=\sqrt{\det G_{AB}}=\frac{g'\left(R\right)g^{2}\left(R\right)\sin\theta}{\sqrt{\Gamma\left(R\right)}},\qquad\sqrt{\det g_{ab}}=r^{2}\left(R\right)\sin\theta\,.
\label{eq:square-root-of-determinant}
\end{equation}
The non-vanishing connection coefficients in $\mathcal{B}$ are 
\begin{equation}
\Gamma_{RR}^{R}=\frac{g''}{g'}-\frac{\Gamma'}{2\Gamma},\qquad\Gamma_{\theta\theta}^{R}=-\frac{g\Gamma}{g'},\qquad\Gamma_{\phi\phi}^{R}=-\frac{g\Gamma\sin^{2}(\theta)}{g'},\qquad\Gamma_{R\theta}^{\theta}=\Gamma_{\theta R}^{\theta}=\Gamma_{R\phi}^{\phi}=\Gamma_{\phi R}^{\phi}=\frac{g'}{g}\,,\label{eq:sphere-connection-Gamma}
\end{equation}
and the non-vanishing Christoffel symbols in $\mathcal{S}$ 
\begin{equation}
\gamma_{\theta\theta}^{r}=-r,\quad\gamma_{\phi\phi}^{r}=-r\sin^{2}\theta,\quad\gamma_{r\theta}^{\theta}=\gamma_{\theta r}^{\theta}=\gamma_{r\phi}^{\phi}=\gamma_{\phi r}^{\phi}=1/r.\label{eq:sphere-connection-gamma}
\end{equation}
Finally, the deformation gradient and its inverse (note the index placement, which has to satisfy $\left(F^{-1}\right)_{\;\:a}^{A}F_{\;\:B}^{a}=\delta_{B}^{A}$ or equivalently $F_{\;\:A}^{a}\left(F^{-1}\right)_{\;\:b}^{A}=\delta_{b}^{a}$):
\begin{equation}
F_{\;\:A}^{a}=\left(\begin{array}{ccc}
r'\left(R\right) & 0 & 0\\
0 & 1 & 0\\
0 & 0 & 1
\end{array}\right),\qquad\left(F^{-1}\right)_{\;\:a}^{A}=\left(\begin{array}{ccc}
\frac{1}{r'\left(R\right)} & 0 & 0\\
0 & 1 & 0\\
0 & 0 & 1
\end{array}\right)\,.\label{eq:sphere-F}
\end{equation}
Incompressibility in the metric theory is $J=1$, see (\ref{eq:J-metric-theory}). We have 
\begin{equation}
J=\frac{\sqrt{\det g_{ab}}}{\sqrt{\det G_{AB}}}\det F_{\;\:A}^{a}=\frac{\sqrt{\Gamma}r^{2}}{g'g^{2}}r'=1\,.
\end{equation}
With this, we note the incompressibility condition
\begin{equation}
r'=\frac{g'g^{2}}{\sqrt{\Gamma}r^{2}}\,.\label{eq:incompressibility}
\end{equation}

\subsection{\label{subsec:spheroid-Ricci-and-Einstein}Ricci and Target Stress tensor}

The calculation of the target stress tensor \eqref{eq:target-stress-quadratic} requires the Ricci tensor $R_{AB}$ (\ref{eq:Ricci-tensor})$_{1}$ and the Ricci scalar $\mathcal{R}$ (\ref{eq:Ricci-tensor})$_{2}$. The non-zero components of the Ricci tensor are given by (\ref{eq:Ricci-tensor})$_{1}$:
\begin{equation}
R_{\;\:R}^{R}=-\frac{\Gamma'}{gg'},\qquad R_{\;\:\theta}^{\theta}=R_{\;\:\phi}^{\phi}=-\frac{1}{g^{2}}\left(\frac{g\Gamma'}{2g'}+\Gamma-1\right)\,.
\end{equation}
The Ricci scalar $\mathcal{R}$ (\ref{eq:Ricci-tensor})$_{2}$ is
\begin{equation}
\mathcal{R}=\frac{2}{g^{2}}\left(-\frac{g\Gamma'}{g'}-\Gamma+1\right)\,.\label{eq:Ricci-scalar-sphere-METRIC}
\end{equation}
As expected, in the compatible case ($\Gamma=1$) all of the above curvature-related quantities (Ricci tensor, scalar) vanish. 

Finally, we have all the ingredients to calculate the target stress $(T^{*})_{\;\:B}^{A}$ \eqref{eq:target-stress-quadratic}:
\begin{equation}
\begin{alignedat}{1}\left(T^{*}\right)_{\;\:R}^{R} & =\frac{2\lambda\left(g'\left(\Gamma'^{2}-4\Gamma\Gamma''\right)+4\Gamma g''\Gamma'+g'^{3}(\mathcal{R}^{*}-\mathcal{R}^{*}\Gamma)\right)}{g^{2}g'^{3}}+\frac{2\lambda(\Gamma-1)(7\Gamma+1)}{g^{4}}-\frac{\lambda\mathcal{R}^{*}{}^{2}}{2}+W^{*}\\
\left(T^{*}\right)_{\;\:\theta}^{\theta}=\left(T^{*}\right)_{\;\:\phi}^{\phi}= & \frac{-4\lambda g^{3}g'^{2}\left(\Gamma'\Gamma''+2\Gamma\Gamma^{(3)}\right)+4\lambda g^{3}g'\left(g''\left(\Gamma'^{2}+6\Gamma\Gamma''\right)+2g^{(3)}\Gamma\Gamma'\right)-24\lambda g^{3}\Gamma g''^{2}\Gamma'}{2g^{4}g'^{5}}\\
 & \phantom{=}-\frac{+g'^{5}\left(g^{4}\left(2W^{*}-\lambda\mathcal{R}^{*}{}^{2}\right)+4\lambda(\Gamma(6-7\Gamma)+1)\right)-2\lambda gg'^{4}\Gamma'\left(\mathcal{R}^{*}g^{2}-14\Gamma+6\right)}{2g^{4}g'^{5}}
  \label{eq:target-stress-sphere-METRIC}
\end{alignedat}
\end{equation}

\subsection{\label{subsec:spheroid-mechanics}Mechanics: Force balance, Eshelby stress}

For a neo-Hookean incompressible material, the strain-energy density and stress-strain relationship are
\begin{equation}
W=\mu\left(C_{\;\:A}^{A}-3\right),\qquad\sigma_{\;\:b}^{a}=\mu B_{\;\:b}^{a}-p\delta_{b}^{a}\,,\label{eq:neo-hookean-incompressible-W-and-sigma}
\end{equation}
where $C_{AB}$ and $B^{ab}$ are the right and left Cauchy-Green strain tensors (\ref{eq:Cauchy-green-tensors}). Firstly, with the right Cauchy-Green tensor being defined as $C_{AB}=\left(F^{T}\right)_{A}^{\;\:a}g_{ab}F_{\;\:B}^{b}$, its trace is 
\begin{align}
C_{\;\:A}^{A} & =G^{AB}C_{AB}=G^{RR}\left(F^{T}\right)_{R}^{\;\:r}g_{rr}F_{\;\:R}^{r}+G^{\Theta\Theta}\left(F^{T}\right)_{\Theta}^{\;\:\theta}g_{\theta\theta}F_{\;\:\Theta}^{\theta}+G^{\Phi\Phi}\left(F^{T}\right)_{\Phi}^{\;\:\phi}g_{\phi\phi}F_{\;\:\Phi}^{\phi}=\frac{g^{4}}{r^{4}}+2\frac{r^{2}}{g^{2}}\,.\label{eq:I1-metric}
\end{align}
This allows us to compute the strain-energy density (\ref{eq:neo-hookean-incompressible-W-and-sigma})
\begin{equation}
W=\mu\left(C_{\;\:A}^{A}-3\right)=\frac{\mu}{2}\left(\frac{g^{4}}{r^{4}}+2\frac{r^{2}}{g^{2}}-3\right)\,.
\label{eq:elastic-energy}
\end{equation}
For the neo-Hookean constitutive relationship, the main challenge lies in computing the left Cauchy-Green strain tensor, which is defined as $B^{bk}=F_{\;\:B}^{b}G^{BA}\left(F^{T}\right)_{A}^{\;\:k}$ (\ref{eq:Cauchy-green-tensors}). Lowering the second index with the spatial metric, we have
\begin{equation}
B_{\;\:r}^{r}=r'^{2}\frac{\Gamma}{g'^{2}},\qquad B_{\;\:\theta}^{\theta}=B_{\;\:\phi}^{\phi}=\frac{r^{2}}{g^{2}}\,.\label{eq:Bab-sphere}
\end{equation}
The Cauchy stress for an incompressible neo-Hookean material is $\sigma_{\;\:b}^{a}=\mu B_{\;\:b}^{a}-p\delta_{b}^{a}$, and in components: 
\begin{equation}
\sigma_{\;\:r}^{r}=\mu\frac{g^{4}}{r^{4}}-p,\qquad\sigma_{\;\:\theta}^{\theta}=\sigma_{\;\:\phi}^{\phi}=\mu\frac{r^{2}}{g^{2}}-p\,.\label{eq:cauchy-stress-components}
\end{equation}
Force balance is given by the divergence of the Cauchy stress tensor. The divergence is obtained by contracting over the first index (by simply setting $c=a$ in (\ref{eq:covariant-derivative-definition})). The divergence of a $\left(1,1\right)$ tensor is 
\begin{align}
\nabla_{a}\sigma_{\;\:b}^{a} & =\partial_{a}\sigma_{\;\:b}^{a}+\gamma_{ac}^{a}\sigma_{\;\:b}^{c}-\gamma_{ab}^{c}\sigma_{\;\:c}^{a}\,.
\end{align}
The only non-trivial component corresponds to $b=r$. By using the connection coefficients (\ref{eq:sphere-connection-gamma}) and by taking into account $\sigma_{\;\:\theta}^{\theta}=\sigma_{\;\:\phi}^{\phi}$, (see (\ref{eq:cauchy-stress-components})), we obtain the familiar formula
\begin{equation}
\nabla_{a}\sigma_{\;\:r}^{a}=\partial_{r}\sigma_{\;\:r}^{r}+\frac{2}{r}\left(\sigma_{\;\:r}^{r}-\sigma_{\;\:\theta}^{\theta}\right)=0\,.\label{eq:Cauchy-force-balance-sphere}
\end{equation}
Using the linear momentum balance (\ref{eq:Cauchy-force-balance-sphere}), together with the constitutive relationship (\ref{eq:cauchy-stress-components}), we obtain:
\begin{equation}
\sqrt{\Gamma} r^{7} \sigma_{\;\:r}^{r}{'}=2\mu g'\left(r^{6}-g^{6}\right)\,.
\label{eq:force-balance}
\end{equation}
Next, our goal is to calculate the Eshelby stress $T_{\;\:B}^{C}=W\delta_{B}^{C}-J\left(F^{-1}\right)_{\;\:b}^{C}\sigma_{\;\:d}^{b}F_{\;\:B}^{d}$, see (\ref{eq:Eshelby-definition}): 
\begin{equation}
T_{\;\:R}^{R}=W-2\mu\frac{g^{4}}{r^{4}}+p,\qquad T_{\;\:\theta}^{\theta}=T_{\;\:\phi}^{\phi}=W-2\mu\frac{r^{2}}{g^{2}}+p\,.\label{eq:Eshelby-stress-sphere-METRIC}
\end{equation}

\subsection{\label{subsec:spheroid-full-problem}Full growth evolution problem}

We now have all the ingredients to write down the full set of equations for an incompressible sphere at mechanical equilibrium that follows a growth law \eqref{eq:generic-growth-law}. The force balance, $\nabla_{a}\sigma_{\;\:b}^{a}=0$, and the incompressibility condition $J=1$, have previously been stated. On top of that, the growth law $\dot{G}_{AB}=k\left(T_{AB}^{*}-T_{AB}\right)$ can now be obtained from previously stated components of the Eshelby stress $T_{\;\:B}^{A}$ and target stress $\left(T^{*}\right)_{\;\:B}^{A}$. The following set of equations contains the fully non-linear equations, boundary conditions, initial conditions, and references to which results from the previous text were used:
\begin{equation}
\begin{alignedat}{3}\text{Eq. }\text{\eqref{eq:force-balance}} & \text{: force balance} & 2\mu\left(r^{6}-g^{6}\right)g{'} & =r^{7}\sqrt{\Gamma}\sigma_{\;\:r}^{r}{'} & \sigma_{\;\:r}^{r}\left(B,t\right)=0\\
\text{Eq. }\text{\eqref{eq:incompressibility}} & \text{: incompressibility}\quad & g^{2}g' & =r^{2}\sqrt{\Gamma}r{'} & r\left(0,t\right)=0\\
\text{Eq. }\text{\eqref{eq:target-stress-sphere-METRIC}, \eqref{eq:Eshelby-stress-sphere-METRIC}} & \text{: growth law}\quad & k\left(T_{RR}^{*}-T_{RR}\right) & =\dot{G}_{RR} & g\left(0,t\right)=0,\quad\Gamma\left(0,t\right)=1\\
\text{Eq. }\text{\eqref{eq:target-stress-sphere-METRIC}, \eqref{eq:Eshelby-stress-sphere-METRIC}} & \text{\text{: growth law}\quad\ } & k\left(T_{\theta\theta}^{*}-T_{\theta\theta}\right) & =\dot{G}_{\theta\theta}\qquad & \mathcal{R}\left(B,t\right)=\mathcal{R}^{*},\quad\partial_{R}\mathcal{R}\left(B,t\right)=0\\
 &  &  &  & \Gamma\left(R,0\right)=\Gamma^{0}\left(R\right),\quad g\left(R,0\right)=g^{0}\left(R\right)
\end{alignedat}
\label{eq:EH-sphere-full-system-with-redundancy}
\end{equation}
Instead of addressing this system as a dynamical problem, we will focus in the following section in the equilibrium scenario ($\dot{G}_{AB}=0$) in which case $T_{AB}=T^*_{AB}$. As we will see, this greatly simplifies the analysis while preserving the physical picture. 

The energy functional (\ref{eq:energy-functional}), using the expression \eqref{eq:elastic-energy} for the elastic energy $W$, the expression \eqref{eq:Ricci-scalar-sphere-METRIC} for the curvature scalar $\mathcal{R}$, and \eqref{eq:square-root-of-determinant} for the square root of the metric determinant $\sqrt{G}$, takes the following form  in the spherically symmetric case 
\begin{equation}
S=4\pi\int_{R=0}^{B}\frac{g'\left(R\right)g^{2}}{\sqrt{\Gamma}}\left\{ \frac{\mu}{2}\left(\frac{g^{4}}{r^{4}}+2\frac{r^{2}}{g^{2}}-3\right)-W^{*}+\frac{\lambda}{2}\left[\frac{2}{g^{2}}\left(-\frac{g\Gamma'}{g'}-\Gamma+1\right)-\mathcal{R}^{*}\right]^{2}\right\} \,dR\,.
\label{eq:functional-explicit-dR}
\end{equation}

\section{\label{sec:gauge-fix}Simplifying the equations through a gauge fix}

\subsection{\label{sec:diffeomorphism-invariance}Dynamical vs. equilibrium views; diffeomorphism invariance}

Mechanical homeostasis is achieved when the tissue stress $T_{AB}$ has reached the homeostatic state $T^*_{AB}$ \eqref{eq:mechanical-homeostasis-1}, the tissue is in a "preferential" state in which growth stops ($\dot{G}_{AB}=0$). An interesting feature of the equilibrium equations 
\begin{equation}
    T_{AB}=T_{AB}^{*}
\end{equation}
is their diffeomorphism invariance. Focusing on the material manifold $\mathcal{B}$, a diffeomorphism is a change of coordinates $X^{A}\rightarrow X^{A'}\left(X^{A}\right)$ under which a scalar field $\phi$, vector field $V^{A}$ and a second order tensor field $T_{AB}$ transform as follows:
\begin{equation}
\phi(X^{A'})=\phi(X^{A}),\qquad V^{A'}=\frac{\partial X^{A'}}{\partial X^{A}}V^{A},\qquad T_{A'B'}=\frac{\partial X^{A}}{\partial X^{A'}}\frac{\partial X^{B}}{\partial X^{B'}}T_{AB}\,.
\label{eq:diffeomorphism-invariance}
\end{equation}
A diffeomorphism invariant scalar field $\phi$ does not change its value under a diffeomorphism; a second order tensor field is defined precisely by the transformation behaviour of $T$. 

Due to the transformation properties of $F_{\;\:A}^{a}=\partial\varphi^{a}/\partial X^{A}$ and the metric tensor $G_{AB}$, the Eshelby stress $T_{AB}$ is indeed consistent with the transformation behaviour \eqref{eq:diffeomorphism-invariance} and thus diffeomorphism invariant. Similarly, due to the transformation properties of Ricci curvature and covariant derivatives, the homeostatic stress $T_{AB}^{*}$ is consistent with \eqref{eq:diffeomorphism-invariance}. Thus, the equations $T_{AB}=T_{AB}^{*}$ are diffeomorphism invariant or (as it is called especially in the context of General Relativity) generally covariant.

A consequence of diffeomorphism invariance is certain fields may be eliminated by finding a coordinate transformation that turns these fields into coordinates, which is  a\textit{ gauge fix}.

To illustrate this concept, let us consider the growing sphere. We introduce a coordinate transformation. Instead of expressing all quantities in terms of the reference coordinate $R$, we express them in terms of the compatible part of the growth tensor, $g$, via the coordinate transformation
\begin{equation}
R\rightarrow g,\qquad r'\left(R\right)\rightarrow g'\left(R\right)r'\left(g\right),\qquad\sigma_{\;\:r}^{r}{'}\left(R\right)\rightarrow g'\left(R\right)\sigma_{\;\:r}^{r}{'}\left(g\right),\qquad\Gamma'\left(R\right)\rightarrow g'\left(R\right)\Gamma'\left(g\right)\,.\label{eq:coordinates-g}
\end{equation}
We introduce the line element, defined as $dS^{2}=G_{AB}dX^{A}dX^{B}$. The line element in old coordinates, which simply restates \eqref{eq:metric-sphere-G} in more compact notation, is 
\begin{equation}
dS^{2}=\frac{g'\left(R\right)^{2}}{\Gamma\left(R\right)}dR^{2}+g^{2}d\Omega^{2}\label{eq:line-element-old-coords}
\end{equation}
where the angular part $d\Omega^{2}=d\Theta^{2}+\sin^{2}\Theta d\Phi^{2}$.  To transform the line element to the new coordinates (\ref{eq:coordinates-g}), we note that $dg=g'\left(R\right)dR$ and thus $dR^{2}=\left[g'\left(R\right)\right]^{-2}dg^{2}$. With this, the previous line element becomes
\begin{equation}
dS^{2}=\frac{1}{\Gamma\left(g\right)}dg^{2}+g^{2}d\Omega^{2}\,.\label{eq:gauge-fixed-metric}
\end{equation}
Compared to \eqref{eq:line-element-old-coords},  we have  performed a \textit{gauge fix}, thus simplifying the metric.

We use the diffeomorphism invariance of the equilibrium equations to our advantage, and from here on study the equilibrium problem
\begin{equation}
    T_{AB}=T_{AB}^{*}\,.
\end{equation}
By performing a gauge fix, we can avoid the extremely complex problem of computing the dynamics of growth and curvature \cite{hamilton1982three} but still extract the essential physical ingredients. Even without knowing the explicit form of $g(R)$ due to the gauge invariance, we will be able to accurately predict the size and energy of the growing spheroid. 

\subsection{Gauge fix}
From here onward, we focus on the equilibrium system given by  $T_{\;\:R}^{R}=\left(T^{*}\right)_{\;\:R}^{R}$ and $T_{\;\:\theta}^{\theta}=\left(T^{*}\right)_{\;\:\theta}^{\theta}$. In the following, it is implicit that $r$ and $\sigma_{\;\:r}^{r}$ are functions of $g$, and the prime denotes derivatives with respect to $g$. With the highest differential order equation $T_{\;\:\theta}^{\theta}=\left(T^{*}\right)_{\;\:\theta}^{\theta}$ eliminated, the fully reduced system is: 
\begin{equation}
\begin{alignedat}{3} & \text{ force balance} & \sigma_{\;\:r}^{r}{'}(g) & =\frac{2\mu\left(r^{6}-g^{6}\right)}{\sqrt{\Gamma}r^{7}} & \sigma_{\;\:r}^{r}\left(g_{B}\right)=0\\
 & \text{incompressibility} & r'(g) & =\frac{g^{2}}{\sqrt{\Gamma}r^{2}} & r\left(0\right)=0\\
 & \text{growth law }\qquad\qquad & \Gamma''(g) & =\frac{-g^{8}\mu-2g^{2}\mu r^{6}+r^{4}g^{4}\left(3\mu-\lambda\mathcal{R}^{*}{}^{2}+2W^{*}\right)}{16g^{2}\lambda\Gamma r^{4}} & \mathcal{R}\left(g_{B}\right)=\mathcal{R}^{*}\\
 & T_{\;\:R}^{R}=\left(T^{*}\right)_{\;\:R}^{R} &  & \phantom{+}+\frac{4\lambda r^{4}\left[g^{2}\Gamma'^{2}+\Gamma\left(-\mathcal{R}^{*}g^{2}+7\Gamma-6\right)\right]}{16g^{2}\lambda\Gamma r^{4}}\qquad\qquad & \mathcal{R}'\left(0\right)=0\\
 &  &  & \phantom{=}+\frac{4r^{4}g^{2}\lambda\mathcal{R}^{*}+2r^{4}g^{4}\sigma_{\;\:r}^{r}-4\lambda r^{4}}{16g^{2}\lambda\Gamma r^{4}} & \mathcal{R}'\left(g_{B}\right)=0
\end{alignedat}
\label{eq:sphere-g-dimensional}
\end{equation}
Note that \eqref{eq:sphere-g-dimensional} is a moving boundary problem: The second-order equation $\Gamma''(g)$ is complemented by three boundary conditions $\mathcal{R}|_{g=g_{B}}=\mathcal{R}^{*}$ and $\mathcal{R}'\left(0\right)=\mathcal{R}'\left(g_{B}\right)=0$. The extra boundary condition allows determining $g_{B}$ as an implicit parameter.

In the new variables, the functional \eqref{eq:functional-explicit-dR} becomes
\begin{equation}
S(g_B)=4\pi\int_{g=0}^{g_{B}}\frac{g^{2}}{\sqrt{\Gamma}}\left\{ \frac{\mu}{2}\left(\frac{g^{4}}{r^{4}}+2\frac{r^{2}}{g^{2}}-3\right)-W^{*}+\frac{\lambda}{2}\left[\frac{2}{g^{2}}\left(-g\Gamma'-\Gamma+1\right)-\mathcal{R}^{*}\right]^{2}\right\} \,dg\,.
\label{eq:functional-explicit-dg}
\end{equation}
We note that the boundary condition $\mathcal{R}'\left(g_{B}\right)=0$ is equivalent to finding the global minimum of $S\left(g_{B}\right)$ by second minimisation. The condition for a stationary state of that functional is $S'\left(g_{B}\right)=0$ which by the fundamental theorem of calculus is
\begin{equation}
W^{*}=W|_{g=g_{B}}\label{eq:boundary-condition-Wstar}
\end{equation}
or equivalently $W^{*}=\frac{\mu}{2}(\frac{g^{4}}{r^{4}}+2\frac{r^{2}}{g^{2}}-3)$ at $g=g_{B}$. Thus, the boundary condition (\ref{eq:boundary-condition-Wstar}) is equivalent to $\mathcal{R}'\left(g_{B}\right)=0$. 

\subsection{Non-dimensionalisation}
Before analysing this system, we make two further modifications. Firstly, the target curvature $\mathcal{R}^{*}$ will be written as 
\begin{equation}
\mathcal{R}^{*}=k\mathcal{R}_{+}^{*},\qquad\mathcal{R}_{+}^{*}>0,\qquad k\in\left\{ -1,0,1\right\} \,,
\end{equation}
where $\mathcal{R}_{+}^{*}$ is a positive value and $k$ absorbs the sign of the curvature. This allows us to scale lengths by a positive curvature value $\mathcal{R}_{+}^{*}$. Secondly, it is  instructive to non-dimensionalise the equations (\ref{eq:sphere-g-dimensional}). The natural length scale is the inverse square root of reference curvature, $\left(\mathcal{R}_{+}^{*}\right)^{-1/2}$. Thus, the following scaling is particularly convenient:
\begin{equation}
\left\{ r,g\right\} =\sqrt{\frac{6}{\mathcal{R}_{+}^{*}}}\left\{ \hat{r},\hat{g}\right\} ,\quad\sigma_{\;\:r}^{r}=\mu\hat{\sigma}_{\;\:\hat{r}}^{\hat{r}},\quad W^{*}=\mu\hat{W}^{*},\quad\lambda=\mu\left(\frac{6}{\mathcal{R}_{+}^{*}}\right)^{2}\hat{\lambda},\quad S=4\pi\mu\left(\frac{6}{\mathcal{R}_{+}^{*}}\right)^{3/2}\hat{S}\,.
\label{eq:EH-final-scaling}
\end{equation}
With this scaling, \eqref{eq:sphere-g-dimensional} becomes
\begin{equation}
\begin{alignedat}{3} & \text{ force balance} & \hat{\sigma}_{\;\:\hat{r}}^{\hat{r}}{'}(\hat{g}) & =\frac{2\left(\hat{r}^{6}-\hat{g}^{6}\right)}{\sqrt{\hat{\Gamma}}\hat{r}^{7}} & \hat{\sigma}_{\;\:r}^{r}\left(\hat{g}_{B}\right)=0\\
 & \text{incompressibility} & \hat{r}'(\hat{g}) & =\frac{\hat{g}^{2}}{\sqrt{\hat{\Gamma}}\hat{r}^{2}} & \hat{r}\left(0\right)=0\\
 & \text{growth law }\qquad\qquad & \hat{\Gamma}''(\hat{g}) & =\frac{1}{16\hat{g}^{2}\hat{\lambda}\hat{\Gamma}\hat{r}^{4}}\Biggl(4\hat{r}^{4}\hat{g}^{2}\hat{\lambda}\hat{\Gamma}'^{2}+\left(3+2\hat{W}^{*}\right)\hat{r}^{4}\hat{g}^{4} & \qquad-\frac{\left(\hat{g}_{B}\hat{\Gamma}'(\hat{g}_{B})+\hat{\Gamma}(\hat{g}_{B})-1\right)}{3\hat{g}_{B}^{2}}=k\\
 & T_{\;\:R}^{R}=\left(T^{*}\right)_{\;\:R}^{R} &  & \phantom{+}-4\hat{r}^{4}\hat{\lambda}\hat{\Gamma}\left(6\hat{g}^{2}k-7\hat{\Gamma}+6\right)-2\hat{g}^{2}\hat{r}^{6}\qquad\qquad & \hat{\Gamma}\left(0\right)=1\\
 &  &  & \phantom{=}-4\hat{r}^{4}\hat{\lambda}\left(1-3\hat{g}^{2}k\right)^{2}+2\hat{r}^{4}\hat{g}^{4}\hat{\sigma}_{\;\:\hat{r}}^{\hat{r}}-\hat{g}^{8}\Biggr) & -\frac{\hat{g}_{B}^{2}\hat{\Gamma}''(\hat{g}_{B})-2\hat{\Gamma}(\hat{g}_{B})+2}{3\hat{g}_{B}^{3}}=0
\end{alignedat}
\label{sphere-g-dimensionless-FULL-SYSTEM}
\end{equation}
The non-dimensional boundary conditions stated above for growth, in this order, correspond to $\mathcal{R}\left(\hat{g}_{B}\right)/\mathcal{R}^{*}_+=k$, $\mathcal{R}'\left(0\right)/\mathcal{R}^{*}=0$ and $\mathcal{R}'\left(\hat{g}_{B}\right)/\mathcal{R}^{*}=0$.\\

Note that this system is not dependent on the reference configuration (material manifold). Instead, curvature (encoded in $\hat{g}\propto \sqrt{\mathcal{R}^*_+}g$) replaces the reference configuration. In other words, the natural length scale of the problem is not the reference length (as is typical in elasticity). Instead, the present gauge reveals the inverse square root of curvature as the natural length scale of the problem.

This pair of coupled highly non-linear first-order ordinary differential equations depends only on \emph{two non-dimensional parameters}, $\hat{\lambda}$ and $\hat{W}^*$. The reference curvature $\mathcal{R}^{*}$ has been scaled out, only its sign $k$ affects the equations. 

In the non-dimensional variables, the energy functional  \eqref{eq:functional-explicit-dg} becomes
\begin{equation}
\hat{S}(\hat{g}_{B})=\int_{g=0}^{\hat{g}_{B}}\frac{\hat{g}^{2}}{\sqrt{\hat{\Gamma}}}\left[\frac{1}{2}\left(\frac{\hat{g}^{4}}{\hat{r}^{4}}+2\frac{\hat{r}^{2}}{\hat{g}^{2}}-3\right)-\hat{W}^{*}+2\hat{\lambda}\left(\frac{-\hat{g}\hat{\Gamma}'-\hat{\Gamma}+1}{\hat{g}^{2}}-3k\right)^{2}\right]\,d\hat{g}\,.
\label{eq:functional-explicit-dgh}\end{equation}

\section{\label{sec:homeostatic-equilibrium}The homeostatic equilibrium for a spheroid}

\begin{figure}[t]
	\centering{ \includegraphics[width=1\textwidth]{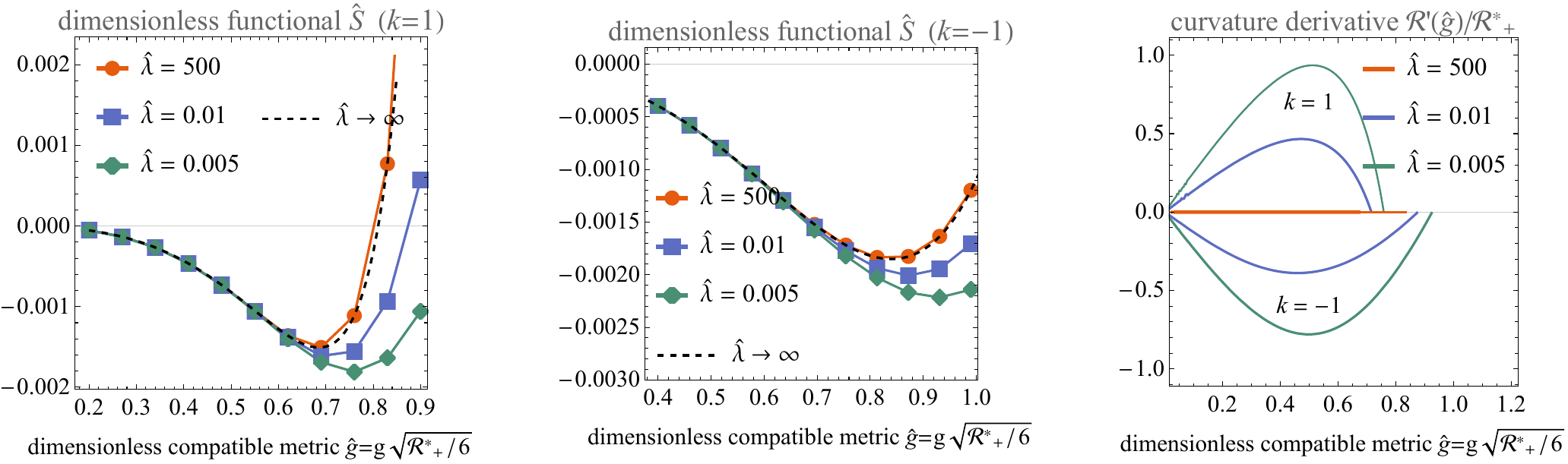}}
	\caption{Dimensionless functional $\hat{S}$ as a function of the dimensionless compatible growth metric $\hat{g}$ for $k=1$ (left panel, positive curvature) and $k=-1$ (right panel, negative curvature). The behaviour is shown for three different values of the dimensionless stiffness parameter $\hat{\lambda}$: 500 (orange circles), 0.01 (blue squares), and 0.005 (green diamonds). The dashed line shows the functional $\hat{S}(\hat{g}_B)$ in the limit $\hat{\lambda} \to \infty$, given by \eqref{eq:functional-uniform-curvature}, corresponding to the regime dominated by the geometric constraint $\mathcal{R}=\mathcal{R}^*$. The plots highlight the impact of the relative weight of $\frac{1}{2}(\mathcal{R}-\mathcal{R}^*)^2$ in the energy functional, showing that the functional has global minima at specific values of $\hat{g}_B$. The derivative of curvature with respect to $\hat{g}$, shown in the right panel, vanishes at $\hat{g}=0$ and $\hat{g}=\hat{g}_B$ where $\hat{g}_B$ is obtained as the minima from the other panels, showing that the boundary condition $W^*=W$ is equivalent to $\partial{\mathcal{R}}/\partial{\hat{g}}=0$ at $\hat{g}=\hat{g}_B$. For all plots, $\hat{W}^*=0.02$.}
	\label{fig:functional}
\end{figure}

We now investigate the  coupled mechanical-geometric problem described by the dimensionless system of equations \eqref{sphere-g-dimensionless-FULL-SYSTEM}. The parameter $\hat{\lambda}$ quantifies the relative strength of the curvature-induced energy penalty compared to elastic energy contributions. By varying $\hat{\lambda}$, we explore the transition from curvature-dominated behaviour ($\hat{\lambda} \gg 1$) to the case where the contribution from elastic energy and curvature penalisation are comparable ($\hat{\lambda} \lesssim 1$), analysing how this mechanical-geometric interplay affects growth, stress distribution, and curvature within the spheroid.

An important result of the proposed theory is that homeostatic equilibrium corresponds, in the spherically symmetric case considered here, to a well-defined equilibrium size for the growing body. This finding, that extends the results of \cite{erlich2024incompatibility}, which were limited to the case where $\hat\lambda\rightarrow\infty$ (uniform curvature), discloses a possibility for growing organisms to control a global feature (size) through local controls alone ($W^*,\mathcal{R}^*$), the latter potentially homogeneous throughout the body, and possessing clear mechanical and geometric interpretation.

\subsection{Numerical solutions}

Firstly, we demonstrate that the boundary conditions $\mathcal{R}\left(\hat{g}_{B}\right)/\mathcal{R}_{+}^{*}=k$, $\mathcal{R}'\left(0\right)=\mathcal{R}'\left(\hat{g}_{B}\right)=0$ are consistent with the boundary condition $\hat{W}^*=\hat{W}(\hat{g}_B)$, which corresponds to the global minimum of the functional \eqref{eq:functional-explicit-dgh}. In Fig. \ref{fig:functional}, we show the functional $\hat{S}\left(\hat{g}_{B}\right)$ for various values of $\hat{g}_{B}$, exhibiting a clear global minimum that increases as $\hat{\lambda}$ decreases, making the spheroid size larger when mechanics and curvature contributions become comparable.  
    
For comparison, we plot the curvature derivatives $\partial\mathcal{R}/\partial\hat{g}$ for the same values of $\hat{\lambda}$. The boundary values for each curvature gradient, $\hat{g}_B$, are those of the global minimum of the respective functional $\hat{S}(\hat{g}_B)$ \eqref{eq:functional-explicit-dgh}. We can clearly see that the boundary conditions $\mathcal{R}'(0)=0$ and $\mathcal{R}'(\hat{g}_B)=0$ are satisfied for all scenarios. This shows that the boundary conditions
\begin{equation}
\mathcal{R}'\left(\hat{g}_{B}\right)=0\qquad\text{and}\qquad W^{*}=W|_{g=g_{B}}\, ,
\end{equation}
where $W^{*}=W|_{g=g_{B}}$, find the global minimum of the functional $\hat{S}(\hat{g}_B)$,  
are equivalent.\\

\begin{figure}[t]
\centering{ \includegraphics[width=1\textwidth]{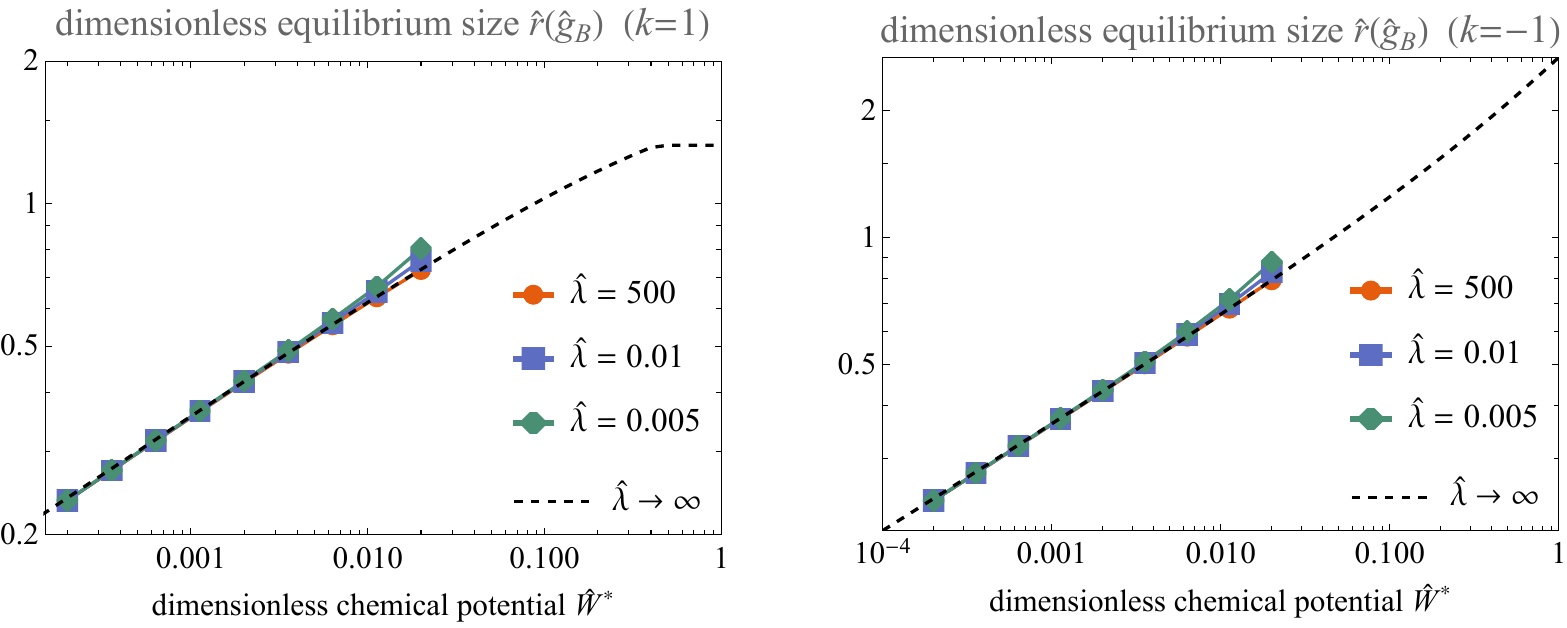}}
\caption{Dimensionless equilibrium size $\hat{r}(\hat{g}_B)$ as a function of the dimensionless compatible growth metric $\hat{g}$, where $k = 1$ (left panel) represents positive curvature and $k = -1$ (right panel) represents negative curvature. The data is shown for different values of the dimensionless stiffness parameter $\hat{\lambda}$: 500 (orange circles), 0.01 (blue squares), and 0.005 (green diamonds). The dashed line corresponds to the theoretical limit $\hat{\lambda} \to \infty$, given by \eqref{eq:functional-uniform-curvature}, corresponding to the regime dominated by the geometric constraint $\mathcal{R}=\mathcal{R}^*$. As the dashed lines indicate, positive Ricci curvature imposes an upper bound on size, whereas negative curvature does not.}
\label{fig:size}
\end{figure}

Next, we investigate how the equilibrium size $\hat{r}(\hat{g}_B)$ is affected by the chemical potential ($\hat{W}^*$) and the strength of mechanical feedback ($\hat{\lambda}$), see Fig. \ref{fig:size}. We overlay the results with the semi-analytical result for $\hat{\lambda}\rightarrow\infty$, which is presented in Section \ref{subsec:spheroid-uniform-curvature}. The effect of $\hat{\lambda}$ on size is modest. For positive curvature ($k=1$), size is bounded, as seen by the asymptote in the figure, whereas for negative curvature ($k=-1$) it is unbounded. \\

\begin{figure}[t]
\centering{ \includegraphics[width=1\textwidth]{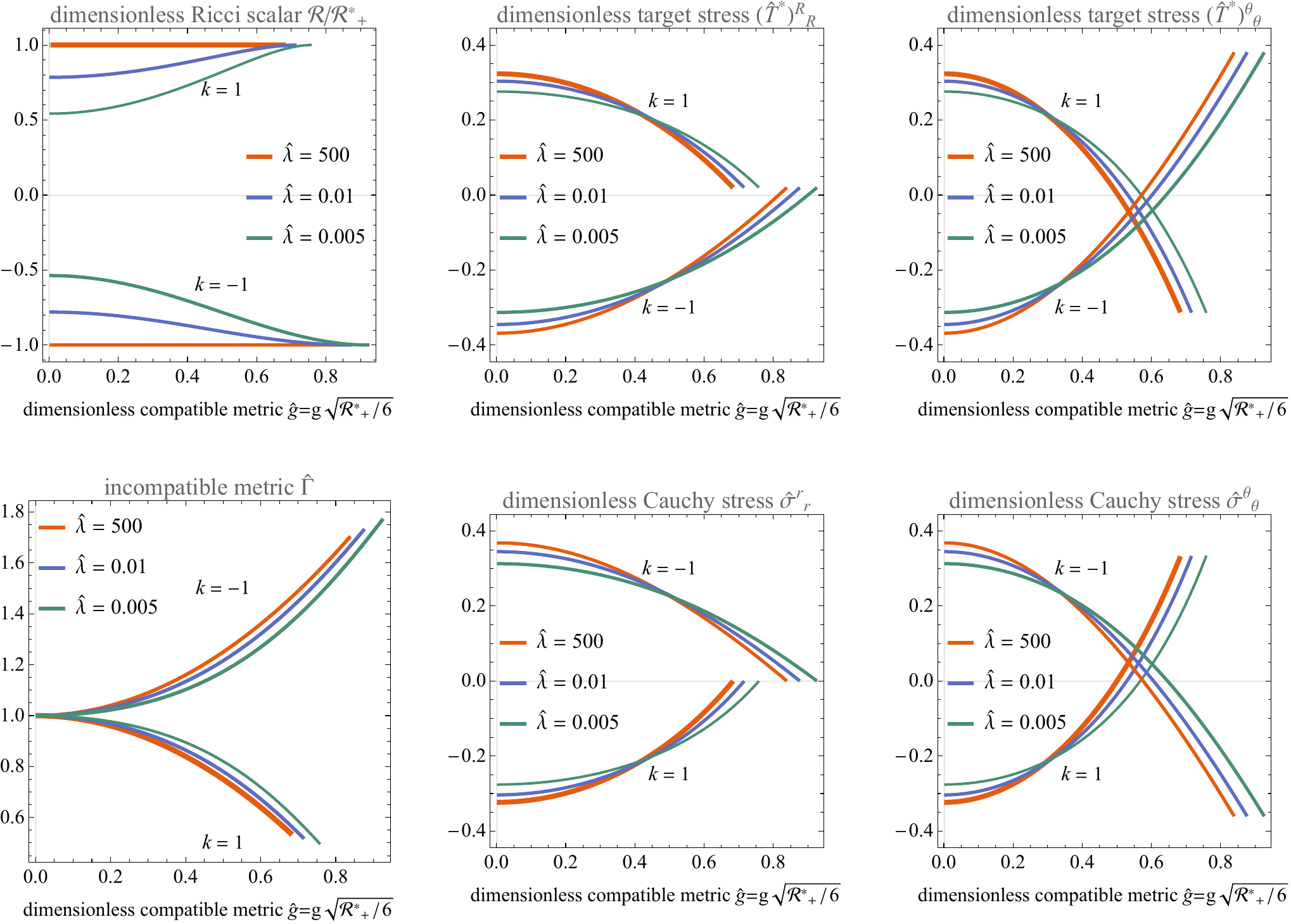}}\caption{\label{fig:fully-coupled-problem-numerics}The scaled Ricci curvature, the incompatible part of the metric, and components of the homeostatic stress and Cauchy stress. All fields are plotted as a function of the dimensionless compatible part of the metric tensor, $\hat{g}$. The fields are computed by solving \eqref{sphere-g-dimensionless-FULL-SYSTEM}. For all plots, $\hat{W}^*=0.02$.}
\end{figure}

Figure \ref{fig:fully-coupled-problem-numerics} presents numerical solutions for both positive (\(k=1\)) and negative (\(k=-1\)) curvature, with varying \(\hat{\lambda}\). The plots show the dimensionless compatible growth metric \(\hat{\Gamma}\), normalized curvature \(\mathcal{R}/\mathcal{R}_+^*\), and the stress components \(\hat{\sigma}_{\;\:r}^{r}\), \(\hat{\sigma}_{\;\:\theta}^{\theta}\), as well as the homeostatic stress $(T^{*})_{\;\:R}^{R}$, $(T^{*})_{\;\:\theta}^{\theta}$, as functions of \(\hat{g}\).

As \(\hat{\lambda}\) decreases, curvature becomes increasingly non-uniform. While \(\mathcal{R} = \mathcal{R}^*\) throughout the spheroid for \(\hat{\lambda} \rightarrow \infty\), finite \(\hat{\lambda}\) leads to increasingly non-uniform curvature the smaller \(\hat{\lambda}\) becomes. The boundary condition $\mathcal{R}=\mathcal{R}^*$ \eqref{eq:sphere-g-dimensional} ensures that curvature matches the reference curvature at the spheroid's periphery. However, the absolute value of curvature decreases towards the centre, regardless of whether $k=1$ or $k=-1$. This behaviour indicates that for smaller values of $\hat{\lambda}$, the distribution of geometric incompatibility becomes more uneven, with most incompatibility being stored near the periphery and less at the centre. Consequently, the smaller $\hat{\lambda}$ gets, the more the interior of the spheroid resembles a uniform flat hydrostatic stress field that is compressive if $k=1$ and tensile if $k=-1$, whereas the peripheral stress always exhibits a steep gradient.

These predictions have important experimental implications. For spheroids with positive curvature ($k=1$) and small values of $\hat{\lambda}$, making a radial cut would result in the spheroid opening more at the periphery and less at the centre. This contrasts with the uniform opening in the curvature-dominated limit ($\hat{\lambda} \rightarrow \infty$), which we have calculated via a finite element simulation in \cite[Fig. 1C]{erlich2024incompatibility}.

\subsection{\label{subsec:spheroid-uniform-curvature}Solutions of uniform curvature}

Here, we consider the limit $\hat{\lambda}\rightarrow\infty$, which enforces that the Ricci curvature scalar $\mathcal{R}$ converges to the target curvature, $\mathcal{R}\rightarrow\mathcal{R}^*$. Studying this limit allows us to derive explicit solutions characterised by uniform curvature. This case was studied in detail in \cite{erlich2024incompatibility}, and here we demonstrate how to recover that theory as a limit case in the present framework.

To this end, we first take the limit $\hat{\lambda}\rightarrow\infty$ of the system \eqref{sphere-g-dimensionless-FULL-SYSTEM}, in which case the growth law simplifies to 
\begin{equation}
\hat{\Gamma}''(\hat{g})=\frac{1}{4}\left(\frac{\hat{\Gamma}'^{2}}{\hat{\Gamma}}-\frac{6\hat{g}^{2}k-7\hat{\Gamma}+6}{\hat{g}^{2}}-\frac{\left(1-3\hat{g}^{2}k\right)^{2}}{\hat{g}^{2}\hat{\Gamma}}\right),\qquad\mathcal{R}/\mathcal{R}_{+}^{*}=k,\qquad\mathcal{R}'\left(0\right)=\mathcal{R}'\left(\hat{g}_{B}\right)=0
\label{eq:ODEs-lambda-inf}
\end{equation}
while the equations for force balance, incompressibility and their boundary conditions are the same as in \eqref{sphere-g-dimensionless-FULL-SYSTEM}. 
As the growth law \eqref{eq:ODEs-lambda-inf} in the limit $\hat{\lambda}\rightarrow\infty$ does not contain Cauchy stress, it can be directly integrated, with the solution
\begin{equation}
\hat{\Gamma}=1-k\hat{g}^{2}\,.
\label{eq:Gamma-uniform-curvature}
\end{equation}
The mechanical and kinematic problems decouple here; the incompressibility condition $\hat{g}^{2}=\sqrt{1-k\hat{g}^{2}}\hat{r}^{2}\hat{r}'$ can be explicitly integrated and gives the current radial variable
\begin{equation}
\hat{r}\left(\hat{g}\right)=\left(\frac{3}{2}\right)^{1/3}\left(\frac{\text{arcsin}\left(\hat{g}\sqrt{k}\right)}{k^{3/2}}-\frac{\hat{g}\sqrt{1-\hat{g}^{2}k}}{k}\right)^{1/3}\,.\label{eq:rhat-constant-curvature}
\end{equation}
This expression was first reported in \cite[Eq. (68)]{erlich2024incompatibility}, here expressed with the scaling \eqref{eq:EH-final-scaling}. In order to find the metric at the boundary, $\hat{g}_B$, we minimse the functional \eqref{eq:functional-explicit-dgh} with the constraint of uniform curvature integrated into it, i.e., we minimse
\begin{equation}
\hat{S}(\hat{g}_{B})=\int_{g=0}^{\hat{g}_{B}}\frac{\hat{g}^{2}}{\sqrt{1-k\hat{g}^{2}}}\left[\frac{1}{2}\left(\frac{\hat{g}^{4}}{\hat{r}^{4}}+2\frac{\hat{r}^{2}}{\hat{g}^{2}}-3\right)-\hat{W}^{*}\right]\,d\hat{g}
\label{eq:functional-uniform-curvature}
\end{equation}
where $\hat{r}$ is given by \eqref{eq:rhat-constant-curvature}. The dashed curve in Fig. \ref{fig:functional} was obtained by plotting \eqref{eq:functional-uniform-curvature} as a function of $\hat{g}_B$, and the dashed curve in Fig. \ref{fig:size} was obtained by plotting the sphere radius \eqref{eq:rhat-constant-curvature} evaluated at $\hat{g}=\hat{g}_B$, where $\hat{g}_B$ was obtained by finding the global minimum of \eqref{eq:functional-uniform-curvature}.

While for negative curvature ($k=-1$), $\hat{r}$ can be arbitrarily large, an interesting feature is that the positive curvature solution ($k=1$) has an upper bound for the spheroid. This bound occurs in the limit of infinite chemical potential (infinite availability of nutrients) $\hat{W}^*\rightarrow\infty$, which corresponds to $\hat{g}_B\rightarrow1$ (see dashed line in Fig.  \ref{fig:size}). Then the spheroid's radius must be below the upper bound for a given target curvature $\mathcal{R}^{*}$:
\begin{equation}
r\leq\frac{3^{5/6}\sqrt[3]{\pi}}{\sqrt[6]{2}}\frac{1}{\sqrt{\mathcal{R}^{*}}}\approx\frac{3.26}{\sqrt{\mathcal{R}^{*}}}\,.\label{eq:upper-bound-size}
\end{equation}

\section{\label{sec:Discussion}Discussion}

In this study, we have presented a novel theoretical framework for modelling morphogenesis by directly integrating growth incompatibility into the system's free energy. This approach overcomes a key limitation of traditional growth models, which rely on an arbitrarily prescribed homeostatic stress tensor \( T_{AB}^{*} \) that depends on unknown parameters and boundary conditions. Our formulation leads to an equilibrium condition of the form \(T_{AB} = T_{AB}^{*}\), where \(T_{AB}^{*}\) now emerges from the variation of the modified energy density and is explicitly related to the curvature of the growth metric tensor \(G_{AB}\). Curvature is a key player in growth because it is the primary source of strain incompatibility and therefore of residual stresses, which in turn are necessary to activate mechanical feedback during growth. While incompatibility, approximated through Ricci curvature, does not by itself distinguish between growth and active contraction processes, the present focus is on mechanical feedback during a growth process that regulates size.

We have shown that penalising deviations from a target curvature \(\mathcal{R}^{*}\) through a term \(\frac{\lambda}{2}(\mathcal{R} - \mathcal{R}^{*})^{2}\) in the free energy leads to a physically transparent expression of the homeostatic Eshelby stress, which is crucial in driving growth and remodelling. In the limit \(\lambda \to \infty\), the curvature \(\mathcal{R}\) matches the target curvature \(\mathcal{R}^{*}\), and the residual stress is fully determined by the prescribed incompatibility, recovering the theory of \cite{erlich2024incompatibility} as a limiting case.

The geometry-based expression of target homeostatic stress constitutes the central outcome of this work. Its clear relationship with geometric and mechanical factors, such as growth incompatibility and chemical potential, highlights the intricacy of the "physiological" state. Our findings elucidate how the internal environment of a growing body can regulate mechanical feedback by addressing the sole source of residual stress—incompatibility—thereby introducing a novel conceptual framework for understanding homeostasis \cite{Cooper2008}. 

The structure of the homeostatic stress that emerges from our formulation offers a valuable alternative to current proposals to explain stress generation in growing systems. In the literature, the choice of homeostatic stress tensor can broadly be categorised based on whether the theory was derived from thermodynamics or not. Theories in which relationships of type \eqref{eq:generic-growth-law} are derived from thermodynamics typically \emph{impose} \(T_{AB}^*\) by hand as a spatially homogeneous anisotropic tensor \cite{ambrosi2007stress,erlich2023mechanical} to allow for residual stress. A way around imposing an anisotropic tensor is to include further internal variables, which has been done in chemo-mechanical and poroelastic models: Here, the choice of an isotropic homeostatic stress tensor of the form \(T^*_{AB}=W^* G_{AB}\), where \(W^*\) is a scalar, is possible and may produce a profile with residual stress because incompatibility arises from the non-uniformity of a concentration field \cite{ambrosi2017solid,xue2016biochemomechanical}. An important point is that in the theories \cite{ambrosi2017solid,xue2016biochemomechanical,ambrosi2007stress}, there is no guarantee of a global minimum of a functional of the type \(S=\int \sqrt{G} W \, \mathrm{d}^3X\) with respect to the growth tensor, meaning that to find the final size of the body, the full growth dynamics—\eqref{eq:generic-growth-law} or its chemo-mechanical or poroelastic equivalent—must be computed.

A second category considers growth dynamics similar in spirit to \eqref{eq:generic-growth-law}, for instance of the form, 
\begin{equation}
\dot{\mathbf{F}}_{g}\mathbf{F}_{g}^{-1}=\mathcal{K}:\left(\boldsymbol{\sigma}-\boldsymbol{\sigma}^{*}\right)\,.\label{eq:old-fashioned-growth-law}
\end{equation}
Here, \(\mathcal{K}\) is a fourth-order tensor of constants, and \(\boldsymbol{\sigma}\) is the Cauchy stress instead of the Eshelby stress. This form is not guaranteed to satisfy the dissipation inequality. In this context, the homeostatic stress equivalent \(\boldsymbol{\sigma}^*\) is typically chosen as an anisotropic constant, and \(\mathcal{K}\) is sometimes selected to couple variables in unconventional ways, such as between radial growth and circumferential stress \cite{taber2001stress,erlich2019homeostatic}. Another category of non-thermodynamical models considers \(\boldsymbol{\sigma}^*\) not as a static field, but instead hypothesises other evolution equations for \(\boldsymbol{\sigma}^*\), allowing the homeostatic stress to evolve and trail behind the Cauchy stress \cite{taber2008theoretical,taber2009towards}. While we acknowledge that our geometry-based formulation excludes many biophysical processes crucial for fully understanding the complexity of biological growth, our study introduces a novel framework to explore how incompatibility can arise at smaller scales within the tissue—an aspect absent in the models referenced above.

Another limitation of the current model lies in the analysis of the "growth action", which necessitates deeper investigations into the existence and uniqueness of minimisers. The example examined in this work, constrained by spherical symmetry, already illustrates the challenges posed by the indeterminacy of the growth tensor in the bulk. This indeterminacy arises from the freedom to apply gauge transformations—mappings that modify the metric without altering the underlying curvature tensor. In the spherical case, this issue was resolved by implementing a judicious gauge-fixing procedure via a coordinate transformation. However, in scenarios with lower symmetry, the question of whether the growth functional admits unique minimisers remains entirely open. Furthermore, a systematic method for addressing superposed gauge transformations in such cases has yet to be established.

A promising avenue of research opened by the current theory is the formulation of an evolution law grounded in \eqref{eq:generic-growth-law}, where the target stress is defined in terms of curvature. This approach bears resemblance to a generalized Ricci flow, a concept originally introduced by Hamilton \cite{hamilton1982three}. Ricci flows have previously been tentatively tried as models for the growth of thin structures like plant leaves \cite{pulwicki2017dynamics,al2018growth}, although a variational or thermodynamical perspective in the context of solid mechanics has not been proposed yet. Ricci flows are renowned for their mathematical intricacies and computational challenges \cite{rubinstein2005visualizing}, especially in higher-dimensional settings. On one hand, the model proposed in this work is expected to recover certain classical results in morphoelasticity with growth dynamics \cite{ambrosi2017solid,erlich2023mechanical}. On the other hand, the analysis of a generalized Ricci flow remains highly nontrivial. Developing robust numerical algorithms will be crucial for investigating the full dynamical behaviour of the theory and validating its theoretical predictions regarding size and stress regulation during growth processes.

One of the most significant implications of our model lies in its biological relevance and experimental testability. As demonstrated in Figure \ref{fig:fully-coupled-problem-numerics}, the spatial distribution of Ricci curvature within the tissue can vary dramatically depending on the value of the coupling parameter \(\lambda\). Specifically, in the limit \(\lambda \to \infty\), the Ricci curvature is uniformly distributed, whereas for small \(\lambda\), the curvature becomes highly inhomogeneous. These variations in curvature distribution directly affect the residual stress profiles within the tissue.

Testing these predictions experimentally would require the development of new protocols capable of precisely measuring the spatial distribution of residual stresses and curvatures in growing tissues. One potential approach involves cutting experiments using micromanipulated blades or lasers to introduce controlled incisions in the tissue and observing the resulting relaxation patterns. Previous studies have shown that elastic tissues with uniform curvature not only open after radial cuts but also exhibit characteristic opening patterns with rounded cut edges \cite{erlich2024incompatibility}. Such patterns offer a method to extract more detailed information than traditional measures like the opening angle, which has been utilised in evaluating residual stresses in multicellular spheroids \cite{guillaume2019characterization, colin2018experimental, stylianopoulos2012causes}. To fully leverage this experimental approach, theoretical work is required to establish optimal cutting protocols that maximise the information gained about the Ricci curvature from a minimal number of cuts. Concurrently, experimental techniques must be refined to perform precise cuts at various spatial locations within small and delicate tissues like embryos and spheroids. This endeavour presents significant challenges due to the technical difficulties in handling and manipulating such biological specimens with the required accuracy.

Another avenue for future research is the microscopic justification of incorporating Ricci curvature into the free energy as a measure of the cost of generating incompatibility within the tissue. In our previous work \cite{erlich2024incompatibility}, we have suggested a potential pathway by defining Ricci curvature at the cellular level using the vertex model, where incompatibility arises from mismatches between a cell's reference area and perimeter. However, this approach needs to be expanded to include incompatibilities stored in the topological connections between cells, reflecting the complex network of interactions in biological tissues.

\section*{Acknowledgements}
We thank Christopher Couzens for valuable discussions on boundary conditions and gauge transformations in general relativity, and for his guidance in computer algebra for differential geometry. We also thank Pierre Recho and Lev Truskinovsky for insightful discussions on growth mechanics. A.E. acknowledges support from the Agence Nationale de Recherche through the JCJC GROWSIZE (ANR-24-CE45-3792).

\section*{Computational codes}

The numerical implementation was done in Mathematica 12.1. The notebooks generating all the figures can be downloaded at: \url{https://github.com/airlich/geometric-nature-of-homeostasis/}.

\appendix

\section{Boundary conditions in the linear (Einstein-Hilbert) and quadratic theory}

We demonstrate in this section  why the linear coupling with curvature, i.e. the Einstein Hilbert action \eqref{eq:einstein-hilbert-action} produces natural boundary conditions for growth which cannot be satisfied. We then continue to show how the quadratic coupling \eqref{eq:quadratic-coupling} fixes this, producing boundary conditions that we are able to satisfy. First, we demonstrate this based on a Lagrangian written explicitly for gauge-fixed metric \eqref{eq:gauge-fixed-metric} which is used throughout the article, by calculating the Euler-Lagrange equations for this specific Lagrangian explicitly rather than relying on the general result \eqref{eq:boundary-subexpressions}. Then, we demonstrate that the boundary conditions obtained in this manual derivation are consistent with the general result \eqref{eq:boundary-subexpressions}.  

The elastic part of the action, $W$, does not contain gradients
of the metric, it does not contribute boundary conditions beyond tractions
as traditional in mechanics \eqref{eq:momentum-balance}. Therefore,
let us focus on the action related to curvature, 
\begin{equation}
S=\int_{\mathcal{B}}\sqrt{G}f\left(\mathcal{R}\right)d^{3}X\,.\label{eq:curvature-action-temporary}
\end{equation}
Consider the gauge fixed metric $dS^{2}=\Gamma^{-1}dg^{2}+g^{2}d\Omega^{2}$
used throughout most of the calculations. For this metric, $\sqrt{G}=g^{2}\Gamma^{-1/2}\sin\theta$.
The Ricci curvature $\mathcal{R}=2\left(1-\Gamma-g\Gamma'\right)/g^{2}$
contains only up to first order derivatives in $g$. This allows us
to write (\ref{eq:curvature-action-temporary}) as
\begin{equation}
S=4\pi\int_{0}^{g_{B}}\mathcal{L}\left(\Gamma,\Gamma'\right)dg
\end{equation}
Taking the variation gives the usual Euler-Lagrange equations in the
bulk plus a boundary term
\begin{equation}
\dot{S}=4\pi\int_{g_{B}}^{g}\left(\frac{\partial\mathcal{L}}{\partial\Gamma}-\frac{d}{dg}\frac{\partial\mathcal{L}}{\partial\Gamma'}\right)\dot{\Gamma}dg+\left[4\pi\frac{\partial\mathcal{L}}{\partial\Gamma'}\dot{\Gamma}\right]_{0}^{g_{B}}
\end{equation}
If we now focus on the boundary term, we can calculate $\partial\mathcal{L}/\partial\Gamma'=-2f'\left(\mathcal{R}\right)g/\sqrt{\Gamma}$.
Thus the variation of the action is $\dot{S}=\dot{S}_{\text{bulk}}+\dot{S}_{\text{bdy}}$
where 
\begin{equation}
\dot{S}_{\text{bdy}}=-8\pi f'\left(\mathcal{R}\right)\dot{\Gamma}\left(g_{B}\right)\,.
\end{equation}
If we assume a General Relativity-like linear coupling $f=\mathcal{R}-\frac{1}{3}\mathcal{R}^{*}$
(see Eq. \eqref{eq:einstein-hilbert-action}), the boundary part of the variation
of the metric is $\dot{S}_{\text{bdy}}\propto g_{B}\Gamma^{-1/2}\left(g_{B}\right)\dot{\Gamma}\left(g_{B}\right)$.
Since we do not know the boundary metric, the
expression $\dot{S}_{\text{bdy}}$ must vanish for all variations
$\dot{\Gamma}\left(g_{B}\right)$, which is achieved only if $g_{B}\Gamma^{-1/2}\left(g_{B}\right)=0$.
This however would imply $\Gamma\left(g_{B}\right)\rightarrow\infty$
which is unphysical. We must therefore discard this Ansatz for $f\left(\mathcal{R}\right)$. 

Conversely, if we assume $f=\frac{1}{2}\left(\mathcal{R}-\mathcal{R}^{*}\right)^{2}$
as in \eqref{eq:quadratic-coupling} and throughout all sphere calculations in the main text, we get 
\begin{equation}
\dot{S}_{\text{bdy}}\propto\left(\mathcal{R}-\mathcal{R}^{*}\right)\dot{\Gamma}\left(g_{B}\right)\,,\label{eq:Sbdy-intermediate}
\end{equation}
and the condition $\mathcal{R}=\mathcal{R}^{*}$ on the boundary satisfies
the natural boundary conditions of the variation. As demonstrated in Fig. \ref{fig:functional} (right panel)
in the main text, determining the yet unknown $g_{B}$ by a second
minimization of the full functional (including elastic energy) provides
a value for $g_{B}$ that is consistent with $\mathcal{R}'\left(g_{B}\right)=0$
at the boundary. \\

This is  consistent with the general result given in \eqref{eq:variation-of-metric-action-BOUNDARY}, \eqref{eq:boundary-subexpressions} of the main text. We will calculate the boundary terms
in \eqref{eq:variation-of-metric-action-BOUNDARY} explicitly to demonstrate this. For our gauge fixed metric
$dS^{2}=\Gamma^{-1}dg^{2}+g^{2}d\Omega^{2}$, the surface normal is
$N=\Gamma^{-1/2}dg$ and the induced metric is $H=g^{2}d\Omega^{2}$.
Then the variation of the boundary term is \eqref{eq:variation-of-metric-action-BOUNDARY} is 
\begin{equation}
\dot{S}_{\text{C}}^{\text{bnd}}=\left[-4\pi g\left(N^{C}M_{C}+N_{D}V^{D}\right)\right]_{g=g_{B}}
\end{equation}
A lengthy evaluation of the terms given in fully by \eqref{eq:boundary-subexpressions}  in the main text reveals that 
\begin{equation}
\dot{S}_{\text{C}}^{\text{bnd}}\propto\left(\frac{(\mathcal{R}-\mathcal{R}^{*})\left(\Gamma-g\Gamma'\right)}{g\Gamma^{3/2}}-\frac{\mathcal{R}'}{\sqrt{\Gamma}}\right)\dot{\Gamma}+\left(\frac{\mathcal{R}-\mathcal{R}^{*}}{\sqrt{\Gamma}}\right)\dot{\Gamma}'\qquad\text{at}\qquad g=g_{B}\,.
\end{equation}
While the terms are somewhat differently arranged than in (\ref{eq:Sbdy-intermediate}),
the vanishing of the $\dot{\Gamma}'$ implies $\mathcal{R}=\mathcal{R}^{*}$
at the boundary, which in turn makes the $\dot{\Gamma}$ vanish if
$\mathcal{R}'\left(g_{B}\right)=0$, so we recover the same  boundary
conditions as before. 

\bigskip

\section*{\textemdash \textemdash \textemdash \textemdash \textemdash \textendash{}}

\bibliographystyle{elsarticle-harv}

\end{document}